\documentclass[11pt]{article}\emergencystretch 3em
\usepackage[utf8]{inputenc}
\usepackage[T1]{fontenc}
\usepackage[right=1in,left=1in,top=1.1in,bottom=1.1in]{geometry}
\usepackage{hyperref}
\hypersetup{colorlinks, citecolor=blue, filecolor=blue, linkcolor=blue, urlcolor=blue}
\usepackage{enumitem}
\usepackage{amsmath,amsthm}
\usepackage{amsfonts}
\usepackage{amssymb}
\usepackage{setspace}

\newtheorem{theorem}{Theorem}[section]
\newtheorem{lemma}[theorem]{Lemma}
\newtheorem{definition}[theorem]{Definition}
\newtheorem{prop}[theorem]{Proposition}
\newtheorem{assumption}[theorem]{Assumption}
\newcommand{\ex}[1]{E \left[ #1 \right]}
\newcommand{\var}[1]{\operatorname{Var} \left[ #1 \right]}

\usepackage[authoryear,round]{natbib}
\newcommand{\textcite}{\citet}
\newcommand{\parencite}{\citep}

\title{Testing Conditional Moment Restrictions in Networks}
\author{Olivia Ding\footnote{Email: \text{olivia.ding@yale.edu}; I am grateful to Yuichi Kitamura, Xiaohong Chen, Tim Christensen, Ed Vytlacil, P. Aronow, Don Andrews, and Max Cytrynbaum for their invaluable advice and encouragement. I thank ChatGPT and Claude for research assistance in literature review and proof checking.}\\
Yale University}

\date{This version: \today\\
(First version: October 1, 2024)}

\begin{document}
\onehalfspacing
\maketitle
\vspace{-2.5em}
\begin{center}
\href{https://drive.google.com/file/d/1-ZHnlvsBgDpzIbvyYavWoiXH4sMbyWwQ/view?usp=sharing}{[Click here for the latest version]}
\end{center}

\begin{abstract}
Parametric models of network formation, such as dyadic logit models, are increasingly popular, but few tools exist to test their specification. This paper
develops a specification test for conditionally independent dyad (CID) models
based on the probabilities that these models assign to subgraphs, the patterns of links among small groups of agents. Given the agents'
characteristics, a parametric CID model determines not only the probability of
each link but also the probability that a group of agents forms a triangle, a
star, or any other subgraph. Under correct specification, the expected
frequency of each subgraph among agents with given characteristics equals its
model-implied probability. This forms the basis of the testable conditional moment
restrictions. The test statistic is the weighted integrated squared distance
between a nonparametric kernel estimate of these probabilities and the
probabilities implied by the estimated parametric model. I derive its
asymptotic distribution under the null hypothesis.

\end{abstract}

\section{Introduction}\label{sec:intro}

Social and economic networks, such as risk-sharing ties among households
\parencite{fafchamps_risk-sharing_2003,fafchamps_risk_2007} and job referral
networks \parencite{schmutte_job_2015}, shape economic outcomes
\parencite{jackson_economic_2017}. A large literature models how they form;
see \textcite{graham_network_2020} and \textcite{de_paula_econometrics_2020} for
overviews. \textit{Conditionally independent dyad} (CID) models, in which links form
independently across pairs of agents given the agents' observed and unobserved
characteristics, are a large class of network formation models. This class
is motivated by the Aldous--Hoover representation of exchangeable random graphs
\parencite{aldous_representations_1981,hoover_relations_1979} and includes
stochastic block models, latent space models, and models with degree
heterogeneity
\parencite{bickel_nonparametric_2009,hoff_latent_2002,graham_econometric_2017,kitamura_estimating_2024}. In parametric CID models, the probability that two
agents link is a known function, such as a logit, of their observed
characteristics and a finite-dimensional parameter; some models add
agent-specific unobserved heterogeneity \parencite{graham_econometric_2017,jochmans_semiparametric_2018,dzemski_empirical_2019,graham_sparse_2024}.
These models are used to measure homophily and degree heterogeneity, to simulate
counterfactual networks, and to serve as building blocks of peer-effects models
with endogenous networks
\parencite{johnsson_estimation_2019,hsieh_specification_2019}. Their conclusions
rest on the parametric specification, yet few tools exist to test it.

This paper develops a specification test for parametric CID models. The test
uses subgraphs: the patterns of links
among small groups of agents, such as triangles and stars. Under correct
specification, the expected frequency of any subgraph among agents with given
characteristics must equal its model-implied probability.
Specifically, for a graph $S$ on $p$ vertices, let $P(S,x)$ be the probability
that the links among $p$ agents with continuously distributed characteristics $x=(x_1,\ldots,x_p)$ form
$S$. Under the parametric model, links are independent given observed
characteristics, and two agents with characteristics $x_i$ and $x_j$ link with probability $e(x_i,x_j;\theta)$. Hence $P(S,x)$ is a known function of these
link probabilities, which I denote by $P(S,x;e_\theta)$, where
$e_\theta:=e(\cdot,\cdot\,;\theta)$. For instance, three agents with characteristics
$(x_1,x_2,x_3)$ form a triangle with probability
$e(x_1,x_2;\theta)e(x_1,x_3;\theta)e(x_2,x_3;\theta)$. The null hypothesis is
that $P(S,x)=P(S,x;e_\theta)$ for some $\theta$ and almost all $x$. I estimate
$P(S,x)$ nonparametrically by the kernel-weighted frequency of $S$ among tuples
of agents whose characteristics are close to $x$
\parencite{ding_nonparametric_2026}. I estimate $\theta$ by the pseudo
composite likelihood estimator of \textcite{graham_sparse_2024}, which yields
the model-implied estimate $P(S,x;e_{\widehat\theta})$. As in kernel-based specification tests
\parencite[e.g.,][]{hardle_comparing_1993}, the
test statistic is the weighted integrated squared difference between the two
estimates,
\[
T_{N,S}
=Nh_N^{d_X/2}\int
\bigl\{\widehat P_N(S,x)
-\mathcal K_{h,N}P(S,\cdot\,;e_{\widehat\theta})(x)\bigr\}^2
\pi(x)\,dx,
\]
where $N$ is the number of agents, $h_N$ is the bandwidth, $d_X$ is the
dimension of each agent's characteristics, $\widehat P_N(S,x)$ is the
nonparametric kernel estimate, and $\pi$ is a weight function. The operator
$\mathcal K_{h,N}$ smooths the parametric estimate with the same kernel
weights, which removes a linear term from the asymptotic expansion of the
statistic. The test applies to any subgraph. Test statistics for several subgraphs
can be combined by taking their maximum or a weighted average; wheels, subgraphs with
one central agent and several arms, are a natural family as their moments
identify stochastic block models
\parencite{bickel_method_2011}. The asymptotic theory in this paper is for a
single subgraph.

The main technical challenge is network dependence. The observations are the
subgraphs induced by $p$-tuples of agents. Two such subgraphs are independent
when their tuples are disjoint but dependent when they share an agent. I first
show that the parametric pseudo composite likelihood estimator is consistent and
asymptotically normal at the $\sqrt N$ rate under this dependence, provided the
model is correctly specified. By the
results of \textcite{ding_nonparametric_2026}, the kernel estimate is uniformly
consistent. I then write the test statistic as a kernel-weighted generalized
U-statistic of order $2p$, derive its Hoeffding decomposition, and isolate its
leading term, a quadratic form. Applying the central limit theorem for
generalized quadratic forms of \textcite{de_jong_central_1987}, I derive the
asymptotic null distribution of the test statistic when its agent-level
projection, the first-order Hoeffding projection onto individual agents, is
nondegenerate.
As with the rate of the kernel estimator in \textcite{ding_nonparametric_2026},
the normalization $Nh_N^{d_X/2}$ depends only on the dimension $d_X$ of a
single agent's characteristics, not on the dimension $pd_X$ of all
characteristics in the subgraph; this alleviates the curse of dimensionality.

\paragraph{Related literature.}
 To the best of my knowledge, this paper provides the
first test of parametric network formation models against nonparametric
alternatives that uses subgraph probabilities conditional on the agents'
characteristics. It contributes to three literatures.

First, it extends specification testing to network data. Specification testing
has a long tradition in econometrics
\parencite{ramsey_tests_1969,hausman_specification_1978,white_consequences_1981,k_newey_generalized_1985},
and consistent tests of conditional moment restrictions against nonparametric
alternatives originate with
\textcite{bierens_consistent_1982,bierens_consistent_1990}. The test here builds
on kernel-based comparisons of parametric and nonparametric fits
\parencite{hardle_comparing_1993,gozalo_consistent_1993,john_xu_zheng_consistent_1996,fan_consistent_1996,fan_consistent_2000};
related tests use series estimators and adaptive procedures
\parencite{hong_consistent_1995,horowitz_adaptive_2001,breunig_adaptive_2024},
integrated conditional moments and empirical processes
\parencite{whang_tests_1993,andrews_conditional_1997,stinchcombe_consistent_1998,chen_consistent_1999,escanciano_consistent_2006},
or empirical likelihood \parencite{tripathi_testing_2003}. These tests were
developed for independent or weakly dependent observations
\parencite{de_jong_bierens_1996}, and recent work extends them to spatial
autoregressive models \parencite{lee_consistent_2020} and to regressors whose
distributions have singular components
\parencite{kankanala_kernel-weighted_2023}. I extend the approach to
observations that are overlapping tuples of agents in a single network.

Second, it contributes to the econometrics of network formation. Parametric CID
models are studied by \textcite{graham_econometric_2017},
\textcite{jochmans_semiparametric_2018}, and \textcite{dzemski_empirical_2019},
who allow for degree heterogeneity, and by \textcite{graham_sparse_2024}, who
studies logistic regression under sparse network asymptotics and possible
misspecification. \textcite{kitamura_estimating_2024} estimate stochastic block
models with covariates, and \textcite{gao_logical_2023} and
\textcite{bonhomme_functional_2023} derive restrictions on model parameters that
are robust to the form of unobserved heterogeneity. Models with strategic
interaction in link formation, which are not CID models, are studied by
\textcite{mele_structural_2017}, \textcite{de_paula_identifying_2018},
\textcite{sheng_structural_2020}, and \textcite{menzel_strategic_2022}, among
others. Existing tests in this literature address specific questions:
\textcite{pelican_optimal_2020} and \textcite{graham_testing_2019} test for
strategic interaction, \textcite{dzemski_empirical_2019} compares observed and
predicted transitivity, and \textcite{auerbach_testing_2022} tests whether two
networks are drawn from the same random graph model. The test proposed here instead checks the parametric specification of link
probabilities against nonparametric alternatives.

Third, it builds on the statistical literature on exchangeable random graphs and
graph limits
\parencite{aldous_representations_1981,hoover_relations_1979,bickel_nonparametric_2009,lovasz_large_2012}.
Subgraph counts have long been used to describe network structure
\parencite{holland_local_1976}, and \textcite{chandrasekhar_network_2025}
model network formation directly in terms of subgraphs rather than links.
\textcite{bickel_method_2011} use subgraph counts to
estimate graphon models by the method of moments, and
\textcite{bhattacharyya_subsampling_2015}, \textcite{green_shalizi_2022}, and
\textcite{lin_lunde_sarkar_2020} develop resampling methods for them.
Goodness-of-fit procedures include simulation-based diagnostics for exponential
random graph models \parencite{hunter_goodness_2008}, tests for the number of
communities in stochastic block models
\parencite{bickel_hypothesis_2016,lei_goodness_2016}, and tests based on small
subgraph probabilities \parencite{gao_testing_nodate}. These procedures use
unconditional subgraph counts or target particular model classes. The test here
instead conditions on the agents' characteristics, so it allows for covariates
and a parametric null. The analysis uses the Hoeffding decomposition of
generalized U-statistics, whose asymptotic distributions in random graphs are
studied by \textcite{janson_asymptotic_1991}. It also relates to work on dyadic
data
\parencite{graham_kernel_2019,graham_minimax_2021,cattaneo_uniform_2023} and on
exchangeable arrays and multiway clustering
\parencite{tabord-meehan_inference_2019,davezies_empirical_2021,menzel_bootstrap_2021,chiang_inference_2021}.
That work mostly concerns statistics indexed by pairs of agents, whereas the
statistic here involves tuples of $p$ agents; for these, the analysis builds on
the uniform convergence results of \textcite{ding_nonparametric_2026} for the
kernel estimator.

\paragraph{Outline.}
Section~\ref{sec:model} introduces the CID model. Section~\ref{sec:moments}
derives the subgraph probabilities and their conditional moment restrictions,
including explicit expressions for wheels. Section~\ref{sec:test} states the
null hypothesis, defines the test statistic, establishes the properties of
the parametric estimator, and states those of the nonparametric estimator. Section~\ref{sec:test-stat-asymptotics}
derives the asymptotic null distribution of the test statistic.
Section~\ref{sec:conclusion} concludes.

\section{Model}\label{sec:model}
I first introduce some graph notation commonly used in studies of networks. Let $G_N = (\mathcal{V}(G_N), \mathcal{E}(G_N))$ be an undirected network (graph) without self-loops on $N$ vertices, or agents. The \textit{vertex set} $\mathcal{V}(G_N)=\{1, \ldots, N\}$ denotes the set of all agents in the network. The \textit{edge list} $\mathcal{E}(G_N)=\{\{i,j\},\{k,l\},\ldots\}$ records the unordered pairs of agents that are linked. I represent $G_N$ by its $N \times N$ adjacency matrix ${D}=\left[D_{i j}\right]_{i, j \in \mathcal{V}(G_N)}$, whose $(i,j)$ element is $1$ if $\{i,j\}\in \mathcal{E}(G_N)$ and $0$ otherwise. See \textcite{graham_network_2020} for details.

Let $X_i\in \mathbb{R}^{d_X}$ be the observed characteristics with support $\mathcal{X}$ and $U_i\in \mathbb{R}$ the unobserved characteristics of agent $i$. Assume that the pairs $\{(X_i,U_i)\}_{i=1}^N$ are independent and identically distributed across agents. Assume also that $U_i$ is independent of $X_i$ and has a continuous distribution with density $f_U$, and that $X_i$ has a continuous distribution. Define $X(G_N) = (X_1, X_2,...,X_N)'\in \mathbb{R}^{Nd_X}$ and $U(G_N) = (U_1, U_2, ..., U_N)'\in \mathbb{R}^{N}$.
I consider the \textit{conditionally independent dyad (CID) model} with covariates and unobserved heterogeneity, in which
$$D_{ij} |X_i, X_j, U_i, U_j \sim \text{Bernoulli}( h(X_i, X_j, U_i,U_j) )$$
independently for every dyad $\{i,j\}$ with $i<j$, where $h(X_i,X_j,U_i,U_j)$ is the probability that $i$ and $j$ form a link. This function $h$ is called the \textit{graphon}; it is measurable, takes values in $[0,1]$, and satisfies $h(x,y,u,v)=h(y,x,v,u)$ for undirected networks. CID models induce a particular dependence structure across the rows and columns of the adjacency matrix: dyads that share an agent are dependent. The model is restrictive in some respects (for example, it rules out strategic interaction among agents), but it is general enough to cover many popular network models, such as stochastic block models, latent space models, and models with degree heterogeneity.

The CID model is a nonparametric model for large networks. It is motivated by the Aldous--Hoover representation theorem for exchangeable random graphs \parencite{aldous_representations_1981,hoover_relations_1979}, under which links in the limiting network form independently conditional on agent-specific variables, here the observed characteristics $X$ and the latent variables $U$. \textcite{chatterjee_estimating_2013} show that dense exponential random graph models are asymptotically of this form.

Under the model, the probability of observing a network $d$ conditional on $X=x$ and $U=u$ is
$$\Pr(D = d |X = x,U = u) = \prod _{i<j} h(x_i,x_j, u_i, u_j)^{d_{ij}} (1-h(x_i,x_j,u_i, u_j))^{1- {d_{ij}}}. $$

Integrating out the unobserved heterogeneity, whose distribution $F$ has density $f_U$, gives the probability of observing $d$ conditional on $X=x$:
$$\Pr(D = d|X = x) =  \int \cdots \int \prod _{i<j} h(x_i,x_j, u_i, u_j)^{d_{ij}} (1-h(x_i,x_j,u_i, u_j))^{1- {d_{ij}}}\prod_{i=1}^N f_U(u_i) du_i. $$
Without loss of generality, I apply the probability integral transform to normalize $F$ to the uniform distribution on $[0,1]$, so that
$$\Pr(D = d|X = x) =  \int_0^1 \cdots \int_0^1 \prod _{i<j} h(x_i,x_j, u_i, u_j)^{d_{ij}} (1-h(x_i,x_j,u_i, u_j))^{1- {d_{ij}}} \prod_{i=1}^Ndu_i. $$

 Networks may be dense or sparse. A common way to model sparsity
\parencite{bickel_method_2011} is to write
    $h_N(x_i,x_j,u_i,u_j) = \rho_N\, w(x_i,x_j,u_i,u_j)  \wedge 1,$
where $w\geq0$ is a fixed function with $\int_0^1\int_0^1 w(x_i,x_j,u_i,u_j)\,du_idu_j=1$, so that, up to the truncation at one, $\rho_N$ is the average link probability among agents with covariates $(x_i,x_j)$. In sparse networks, $\rho_N\to0$, so the expected number of links grows more slowly than $N^2$.
This paper studies dense networks, in which $h$ does not depend on $N$; extending the test to sparse networks is left for future work.

\section{Conditional moments using subgraph counts}\label{sec:moments}
This section derives conditional moment restrictions from the frequency with which a given small pattern of links, a subgraph, appears in a network. I first introduce some definitions and notation for subgraphs.

\subsection{Induced subgraphs and density}
\begin{definition}[Graph Isomorphism]
Graphs $R$ and $S$ are isomorphic, written $R\simeq S$, if there is a bijection $\varphi:\mathcal V(R)\to\mathcal V(S)$ such that $\{a,b\}\in\mathcal E(R)$ if and only if $\{\varphi(a),\varphi(b)\}\in\mathcal E(S)$.
\end{definition}
I say that $(S, x)$ is an \textit{induced subgraph of a graph $G = (\mathcal{V} (G), \mathcal{E} (G))$ on $p$ vertices with covariates $x\in \mathbb{R}^{pd_X}$} if
        (i) $\mathcal{V} (S) \subseteq \mathcal{V} (G)$ contains $p$ vertices; (ii) $\mathcal{E} (S) = \mathcal{E} (G) \cap \{\mathcal{V} (S) \times \mathcal{V} (S)\}$; and (iii) the stacked covariates of the vertices in $\mathcal{V}(S)$ equal $x$.
Although $X_i$ is continuous, it is convenient to treat it as discrete when defining the empirical subgraph density; Section~\ref{subsec:nonparametric-moments} handles continuous covariates.
I define the \textit{induced subgraph density of $S$ with covariates $x\in \mathbb{R}^{pd_X} $ in $G_N$} to be
        \begin{equation}\label{eq:defP_N}
\begin{aligned}
            P_N(S, x) &= \frac{\sum_{\mathbf{i}_p \in \mathcal{P}_{p,N}}\mathbf{1} (S \simeq  G_N [\mathbf{i}_p ], X( G_N [\mathbf{i}_p ]) = x)}{\sum_{\mathbf{i}_p \in \mathcal{P}_{p,N}}\mathbf{1} ( X( G_N [\mathbf{i}_p ]) = x)} \\
            &= \frac{\text{\# of $p$-tuples inducing $S$ with covariates $x$ in } G_N}{\text{\# of $p$-tuples with covariates $x$ in } G_N}
\end{aligned}
        \end{equation}
    where $\mathcal{P}_{p,N} := \{(i_1,\dots,i_p): i_a\in[N],\ i_a\neq i_b\text{ for }a\neq b\}$ is the set of ordered $p$-tuples $\mathbf{i}_p=(i_1,\ldots,i_p)$ of distinct vertices in $[N]:=\{1,\dots,N\}$, with cardinality $M_N:=|\mathcal{P}_{p,N}|=N!/(N-p)!$ for $N\geq p$, and $G_N [\mathbf{i}_p ]$ is the subgraph of $G_N$ induced by the vertices in $\mathbf{i}_p$. The covariate vector is stacked in tuple order, $X_{\mathbf i_p}:=(X_{i_1},\ldots,X_{i_p})$, and the labels $1,\ldots,p$ refer to those positions. I also define $\mathcal{C}_{p,N}:=\{\mathbf q_p=(q_1,\dots,q_p):1\le q_1<\cdots<q_p\le N\}$, the set of unordered \(p\)-vertex subsets, each written in increasing order.
For \(\mathbf q_p\in\mathcal C_{p,N}\), I write a permutation of $\mathbf{q}_p$ as $\sigma(\mathbf{q}_p)$ for $\sigma\in \mathfrak S(\mathbf{q}_p)$, where $\mathfrak S(\mathbf{q}_p)$ denotes the set of all
\(p!\) permutations of \(\mathbf q_p\).

The population counterpart of $P_N(S,x)$ is the probability that $p$ agents
with covariates $x$ induce a copy of $S$:
\[
P(S,x):=\Pr\{G_N[(1,\ldots,p)]\simeq S\mid X_{(1,\ldots,p)}=x\}.
\]
To compute it, label the $p$ agents $1,\ldots,p$, so that agent $a$ has
covariate $x_a$, and let $\mathcal E_p:=\{\{a,b\}:1\leq a<b\leq p\}$ be the
set of all pairs among them. For a labelled graph $R$ on $[p]$, the
probability that the links among agents $1,\ldots,p$ are exactly those of $R$
(every edge of $R$ present and every other pair unlinked) is
\[
P_{\mathrm{lab}}(R,x)
:=\int_{[0,1]^p}
\prod_{\{a,b\}\in\mathcal E(R)}h(x_a,x_b,u_a,u_b)
\prod_{\{a,b\}\in\mathcal E_p\setminus\mathcal E(R)}
\{1-h(x_a,x_b,u_a,u_b)\}\,du.
\]
Let $\mathcal O(S)$ be the set of distinct labelled graphs on $[p]$ that are
isomorphic to $S$. The agents induce a copy of $S$ exactly when their links
form one of the graphs in $\mathcal O(S)$. These events are disjoint, so
\[
P(S,x)=\sum_{R\in\mathcal O(S)}P_{\mathrm{lab}}(R,x).
\]
For example, if $S$ is a path with two edges ($p=3$), $\mathcal O(S)$
contains three graphs, one for each choice of the middle vertex.

For subgraphs that are not necessarily induced, define the labelled
\textit{injective homomorphism density}
\[
Q(S,x):=\int_{[0,1]^p}
\prod_{\{a,b\}\in\mathcal E(S)}h(x_a,x_b,u_a,u_b)\,du.
\]
$Q(S,x)$ is the probability that every edge of $S$, in its labelled position, is present, with no restriction on the other dyads. Note that $P_{\mathrm{lab}}(S,x)$ and $Q(S,x)$ are closely related:
\begin{align*}
P_{\mathrm{lab}}(S,x)
&=\sum_{A\subseteq\mathcal E_p\setminus\mathcal E(S)}
(-1)^{|A|}Q(S\cup A,x),\\
Q(S,x)
&=\sum_{\mathcal E(S)\subseteq\mathcal E(R)\subseteq\mathcal E_p}
P_{\mathrm{lab}}(R,x),
\end{align*}
where $S\cup A$ is the labelled graph with edge set $\mathcal E(S)\cup A$.
These are the covariate-conditional versions of the identities
underlying Proposition~1 in \textcite{bickel_method_2011}. The isomorphism probability $P(S,x)$ then follows by summing $P_{\mathrm{lab}}(R,x)$ over $R\in\mathcal O(S)$, as above.

\subsection{Explicit formula for wheels}
For some subgraphs, the homomorphism density $Q(S,x)$ has an explicit form. Wheels are one example.
    \begin{definition}[Wheels]
     A $(k, l)$-wheel is a graph isomorphic to the graph on $kl+1$ vertices with the $kl$ edges $\{\{1,2\},\ldots,\{k,k+1\}\}\cup\{\{1,k+2\},\ldots,\{2k,2k+1\}\}\cup\cdots\cup\{\{1,(l-1)k+2\},\ldots,\{lk,lk+1\}\}$.
    \end{definition}
    In other words, a wheel consists of a central vertex $1$ and $l$ arms attached to it, each a path of $k$ edges. I consider only $k \geq 2$; for $k=1$, the wheel reduces to a star.
For fixed covariates $x,y$, define the operator
\[
[T_{x,y}f](u):=\int_0^1h(x,y,u,v)f(v)\,dv.
\]
Label the noncentral vertices on arm $m$ by $v_{m,1},\ldots,v_{m,k}$,
and write $x_{m,r}$ and $u_{m,r}$ for their covariates and latent variables.
Set $x_{m,0}=x_1$ and $u_{m,0}=u$, the shared center. Define the conditional
arm moment
\[
A_m(u;x):=\int_{[0,1]^k}
\prod_{r=1}^k h(x_{m,r-1},x_{m,r},u_{m,r-1},u_{m,r})
\,du_{m,1}\cdots du_{m,k}.
\]
Independence of the noncentral vertex variables across arms, conditional on
the center, gives
\begin{align*}
Q(S,x)&=\int_0^1\prod_{m=1}^l A_m(u;x)\,du,\\
A_m(u;x)&=\left[
T_{x_1,x_{m,1}}T_{x_{m,1},x_{m,2}}\cdots
T_{x_{m,k-1},x_{m,k}}\mathbf 1\right](u).
\end{align*}
The wheel moment thus reduces to a one-dimensional integral over the center of
products of arm moments, each obtained by iterating the operators $T_{x,y}$.
The induced probability $P(S,x)$ of a wheel then follows.
Wheels are also informative about the model: Theorem~1 of \textcite{bickel_method_2011} shows that the parameters of a stochastic block model are generically identified by finitely many wheel moments.

\subsection{Network moments using subgraph counts}\label{subsection:Network-moments-using-subgraph-counts}
Write $\mathcal F_{X,N}:=\sigma(X_1,\ldots,X_N)$ and, for each ordered tuple,
define the induced-isomorphism and labelled homomorphism responses
\[
Y_{\mathbf i_p,S}:=\mathbf 1\{G_N[\mathbf i_p]\simeq S\},
\qquad
Z_{\mathbf i_p,S}:=\prod_{\{a,b\}\in\mathcal E(S)}D_{i_ai_b}.
\]
Under the CID model and the independence assumptions in Section~\ref{sec:model}, every
ordered tuple satisfies the conditional moment restrictions
\begin{align*}
E[Y_{\mathbf i_p,S}-P(S,X_{\mathbf i_p})\mid\mathcal F_{X,N}]&=0,\\
E[Z_{\mathbf i_p,S}-Q(S,X_{\mathbf i_p})\mid\mathcal F_{X,N}]&=0.
\end{align*}
For discrete covariates, when the common denominator
is positive, the conditional expectation of $P_N(S,x)$ given
$\mathcal F_{X,N}$ is $P(S,x)$. With continuous
covariates, exact matches occur with probability zero, and the tuple-level
restrictions above motivate the kernel estimators below.

\section{Model specification test}\label{sec:test}

I define the integrated graphon function $g:\mathcal{X}\times \mathcal{X} \to [0,1]$ by
$$ g(X_i,X_j) \equiv E[h(X_i,X_j,U_i, U_j)|X_i,X_j]$$ and parametrize it as $g(X_i,X_j)=e(X_i,X_j;\theta)$ for a finite-dimensional $\theta$. A popular choice is the logit model of \textcite{graham_econometric_2017, graham_sparse_2024}, with $\theta=(\alpha,\beta')'$:
\[
e(X_i,X_j;\theta)
=\frac{\exp\{\alpha+(X_i+X_j)'\beta\}}
{1+\exp\{\alpha+(X_i+X_j)'\beta\}}.
\]
More generally, the index can be $R_{ij}'\theta$ for any
symmetric vector of dyad regressors $R_{ij}=R_{ji}$.
I maintain the hypothesis that the network follows a CID model. For a fixed
subgraph $S$ on $p$ vertices, with $X=(X_1,\ldots,X_p)$, I test the null hypothesis
    \begin{align*}
        H_0: \quad &\Pr\{P(S,X) = P (S,X; e_\theta)\} = 1 \text{  for some  } \theta \in \Theta,\\
        H_1: \quad &\Pr\{P(S,X) = P (S,X; e_\theta)\} < 1 \text{  for all  } \theta \in \Theta.
    \end{align*}

\subsection{Test statistics}
The subgraph moments of Section~\ref{sec:moments} are testable implications of the parametric model. I use them to construct a H\"ardle--Mammen-type test of the parametric null against a nonparametric alternative.
For a fixed induced subgraph \(S\), I consider the network moment $P(S,x).$
Under the parametric null indexed by \(\theta\), the corresponding model-implied
network moment for subgraph $S$ with covariates $x$ is
\[
P(S,x;e_{\theta})
=\sum_{R\in\mathcal O(S)}
\prod_{\{a,b\}\in\mathcal E(R)}e(x_a,x_b;\theta)
\prod_{\{a,b\}\in\mathcal E_p\setminus\mathcal E(R)}
\{1-e(x_a,x_b;\theta)\}.
\]

Given an estimator $\widehat{\theta}$, the parametric fit of $P(S,x)$ is
\begin{equation*}
             P(S,x;e_{\widehat\theta})
=\sum_{R\in\mathcal O(S)}
\prod_{\{a,b\}\in\mathcal E(R)}e(x_a,x_b;\widehat\theta)
\prod_{\{a,b\}\in\mathcal E_p\setminus\mathcal E(R)}
\{1-e(x_a,x_b;\widehat\theta)\}.
        \end{equation*}
The nonparametric fit of $P(S,x)$ is the kernel estimator \(\widehat P_N(S,x)\) defined in Section~\ref{subsec:nonparametric-moments}, which also states its regularity conditions.
The test statistic $T_{N,S}$ for subgraph $S$ is the weighted $L^2$-distance between $\widehat P_N(S,x)$ and a smoothed version of $P(S,x;e_{\widehat\theta})$:
        \begin{equation}\label{eq:test_stats}
           T_{N,S}
=
N h_N^{d_X/2}
\int
\left[
\widehat P_N(S,x)
-
\mathcal{K}_{h,N}P(S,\cdot\,;e_{\widehat\theta})(x)
\right]^2
\pi(x)\,dx,
        \end{equation}
 where $\mathcal{K}_{h,N}$ is the smoothing operator defined by $$\mathcal{K}_{h,N} \varphi (x)  = \frac{\sum\limits_{\mathbf{i}_p \in \mathcal{P}_{p,N}} K_h(x - X_{\mathbf{i}_p}) \varphi(X_{\mathbf{i}_p}) }{\sum\limits_{\mathbf{i}_p \in \mathcal{P}_{p,N}} K_h(x - X_{\mathbf{i}_p})},$$ with $K_h(v):=h_N^{-d_W}K(v/h_N)$ for a kernel $K:\mathbb{R}^{d_W}\to\mathbb{R}$ on the stacked covariate space of dimension $d_W:=pd_X$ and a bandwidth $h_N\to0$; Section~\ref{subsec:nonparametric-moments} states the conditions on $K$ and $h_N$.
Note that $E[\widehat P_N(S,x)\mid X_1,\ldots,X_N]=\mathcal{K}_{h,N}P(S,\cdot)(x)$. Applying the same operator to $P(S,x;e_{\widehat\theta})$ makes $T_{N,S}$ asymptotically equivalent to a constant plus a quadratic form; without it, the statistic would also contain a linear term.

To use several subgraphs at once, for example all wheels in a set $\mathcal{S}$, one can take the maximum or a weighted average of the statistics over $\mathcal{S}$, with weights $w(S)\geq0$:
\begin{equation*}
    T_N^{\max} = N h_N^{d_X/2} \max_{S\in \mathcal{S}} \int ( \widehat P_N(S,x)
-
\mathcal{K}_{h,N}P(S,\cdot\,;e_{\widehat\theta})(x))^2 \pi(x) dx
\end{equation*}
or \begin{equation*}
    T_N^{\mathrm{avg}} = N h_N^{d_X/2} \sum_{S\in \mathcal{S}} w(S) \int ( \widehat P_N(S,x)
-
\mathcal{K}_{h,N}P(S,\cdot\,;e_{\widehat\theta})(x))^2 \pi(x) dx.
\end{equation*}
The asymptotic theory in Section~\ref{sec:test-stat-asymptotics} is for a single
subgraph $S$. The null distributions of $T_N^{\max}$ and $T_N^{\mathrm{avg}}$
depend on the joint limit of $(T_{N,S})_{S\in\mathcal S}$, which I do not derive
here.
Now I turn to estimation of the network moments.
\subsection{Parametric estimation of the subgraph moments}
Following \textcite{graham_sparse_2024}, I estimate the parametric model by maximizing a pseudo composite log-likelihood:
    \begin{align*}
        \widehat{\theta} & = \arg \max_{\theta \in \Theta} L_N(\theta) = \arg \max_{\theta \in \Theta} \frac{1}{N(N-1)}\sum_{i=1}^N \sum_{j\neq i}\ell_{ij,N}(\theta),
    \end{align*}
    where the logit kernel is $\ell_{ij,N}(\theta) \equiv (2D_{ij}-1)(R_{ij}'\theta ) - \ln{(1+\exp{ ((2D_{ij}-1)(R_{ij}'\theta)) })}$ and $R_{ij} = (1, \zeta(X_i, X_j))'$ for a known symmetric transformation $\zeta$, that is, $\zeta(x,y)=\zeta(y,x)$.
Write $e_{ij}(\theta):=\Lambda(R_{ij}'\theta)$ with $\Lambda(t):=e^t/(1+e^t)$, so that $e_{ij}(\theta)=e(X_i,X_j;\theta)$, and let $s_{ij}(\theta):=\partial_\theta \ell_{ij,N}(\theta)=R_{ij}\{D_{ij}-e_{ij}(\theta)\}$ denote the dyad score.
I show below that $\widehat{\theta}$ is $\sqrt{N}$-consistent and asymptotically normal, and I plug it into $P(S,x;e_\theta)$ to obtain $P(S,x;e_{\widehat\theta})$.
I impose the following regularity conditions.

\begin{assumption}[Compactness and interior]\label{ass:parametric_compact}
The parameter space $\Theta$ is compact and $\theta_0 \in \operatorname{int}(\Theta)$.
\end{assumption}

\begin{assumption}[Identification, smoothness, and nonsingularity]\label{ass:parametric_population}
Define $\mathcal L(\theta) \equiv \ex{\ell_{12,N}(\theta)}.$
The function $\mathcal L(\theta)$ is uniquely maximized at $\theta_0$, is twice continuously
differentiable on an open neighborhood of $\theta_0$, and
$\mathcal H_0 \equiv \partial^2 \mathcal L(\theta_0)/\partial \theta \partial \theta'$
is nonsingular.
\end{assumption}

\begin{assumption}[Bounded moments]\label{ass:parametric_moment} Assume
\[
\ex{\sup_{\theta \in \Theta}\left\|
\frac{\partial \ell_{12,N}(\theta)}{\partial \theta}
\right\|^2} < \infty,
\qquad
\ex{\sup_{\theta \in \Theta}\left\|
\frac{\partial^2 \ell_{12,N}(\theta)}{\partial \theta \partial \theta'}
\right\|} < \infty.
\]
Moreover, for any $\theta,\bar\theta \in \Theta$,
$\left|
\ell_{12,N}(\theta)-\ell_{12,N}(\bar\theta)
\right|
\leq L_{12}\|\theta-\bar\theta\|$
for some random variable $L_{12}$ satisfying $\ex{L_{12}}<\infty$.
Assume also that there exists \(\delta>0\) such that $\ex{\|R_{12}\|^{4+\delta}}<\infty.$
\end{assumption}

\begin{assumption}[Correct specification]\label{ass:pcle_an} Assume $E[D_{12}\mid X_1,X_2]=e_{12}(\theta_0)$ a.s.
\end{assumption}

\begin{prop}\label{prop:pcle_consistency}
Under Assumptions~\ref{ass:parametric_compact}--\ref{ass:pcle_an}, $\widehat\theta \xrightarrow{p}\theta_0.$
\end{prop}

\begin{proof}
Write
\[
L_N(\theta)
=
\binom{N}{2}^{-1}\sum_{1\le i<j\le N}\bar\ell_{ij}(\theta)
\]
with $\bar\ell_{ij}(\theta):=\frac{1}{2}\bigl\{\ell_{ij,N}(\theta)+\ell_{ji,N}(\theta)\bigr\}$.
For each fixed $\theta\in\Theta$, the dyad summands are identically distributed, and those of vertex-disjoint dyads are independent, under the CID model with i.i.d. vertex variables. They are integrable because
$|\ell_{ij,N}(\theta)|\leq\log 2+\|R_{ij}\|\,\|\theta\|$ and
$E\|R_{12}\|<\infty$. The law of large numbers for U-statistics \parencite{hoeffding_strong_1961} therefore
gives $L_N(\theta)\xrightarrow{p}E[\bar\ell_{12}(\theta)].$
Because \((X_1,X_2)\) and \((X_2,X_1)\) have the same distribution, $E[\bar\ell_{12}(\theta)]
=
\Bigl(E[\ell_{12,N}(\theta)]+E[\ell_{21,N}(\theta)]\Bigr)/2
=
\mathcal L(\theta).$
Next, by Assumption~\ref{ass:parametric_moment}, $|\bar\ell_{12}(\theta)-\bar\ell_{12}(\bar\theta)|
\le \bar L_{12}\|\theta-\bar\theta\|,$ where $\bar L_{12}:=\frac{1}{2}(L_{12}+L_{21})$ satisfies \(E[\bar L_{12}]<\infty\). Since \(\Theta\) is compact, the usual finite-covering argument yields
$\sup_{\theta\in\Theta}|L_N(\theta)-\mathcal L(\theta)|\xrightarrow{p}0.$
Since \(\mathcal L\) is uniquely maximized at \(\theta_0\) by Assumption~\ref{ass:parametric_population}, $\widehat\theta\xrightarrow{p}\theta_0$ by a standard argument \parencite{newey_chapter_1994}.
\end{proof}

\begin{lemma}[Uniform Hessian convergence]\label{lem:hessian_unif}
Under Assumptions~\ref{ass:parametric_compact}--\ref{ass:pcle_an}, $\sup_{\theta\in\Theta}\|\mathcal H_N(\theta)-E\mathcal H_{12}(\theta)\|\xrightarrow{p}0,$
where
\[
\mathcal H_N(\theta):=\partial_{\theta\theta'}^2L_N(\theta)
=
\frac{1}{N(N-1)}\sum_{i\ne j}\mathcal H_{ij}(\theta),
\]
\[
\mathcal H_{ij}(\theta)
:=
\partial_{\theta\theta'}^2\ell_{ij,N}(\theta)
=
-\,e_{ij}(\theta)\{1-e_{ij}(\theta)\}R_{ij}R_{ij}'.
\]
\end{lemma}

\begin{proof}
Direct calculation gives the score, the Hessian, and the derivative of the Hessian of the logit kernel $\ell_{ij,N}(\theta)$:
\[
s_{ij}(\theta)=\partial_\theta \ell_{ij,N}(\theta)
=
R_{ij}\{D_{ij}-e_{ij}(\theta)\},
\]
\[
\mathcal H_{ij}(\theta)
=
\partial_{\theta\theta'}^2\ell_{ij,N}(\theta)
=
-\,e_{ij}(\theta)\{1-e_{ij}(\theta)\}R_{ij}R_{ij}',
\]
and, for each coordinate \(r\in\{1,\dots,\dim(\theta)\}\),
\[
\frac{\partial \mathcal H_{ij}(\theta)}{\partial\theta_r}
=
-\,e_{ij}(\theta)\{1-e_{ij}(\theta)\}\{1-2e_{ij}(\theta)\}
R_{ij,r}\,R_{ij}R_{ij}'.
\]
Since \(0\le e_{ij}(\theta)\le 1\),
$\sup_{\theta\in\Theta}\|\mathcal H_{ij}(\theta)\|
\le \frac{1}{4}\|R_{ij}\|^2,$
and
\[
\sup_{\theta\in\Theta}
\left\|
\frac{\partial \mathcal H_{ij}(\theta)}{\partial\theta_r}
\right\|
\le C\,|R_{ij,r}|\,\|R_{ij}\|^2
\le C\,\|R_{ij}\|^3
\]
for a finite constant \(C\). Therefore, by the mean-value theorem, $\|\mathcal H_{ij}(\theta)-\mathcal H_{ij}(\bar\theta)\|
\le C\,\|R_{ij}\|^3\,\|\theta-\bar\theta\|.$

Averaging over ordered dyads gives
\[
\|\mathcal H_N(\theta)-\mathcal H_N(\bar\theta)\|
\le
\Biggl(
\frac{C}{N(N-1)}\sum_{i\ne j}\|R_{ij}\|^3
\Biggr)\|\theta-\bar\theta\|.
\]
Since \(E\|R_{12}\|^3<\infty\), the law of large numbers for U-statistics \parencite{hoeffding_strong_1961} implies
\[
\frac{1}{N(N-1)}\sum_{i\ne j}\|R_{ij}\|^3=O_p(1),
\]
so \(\{\mathcal H_N(\theta):\theta\in\Theta\}\) is stochastically equicontinuous.
For each fixed \(\theta\in\Theta\),
\[
\mathcal H_N(\theta)
=
\binom{N}{2}^{-1}\sum_{1\le i<j\le N}\bar{\mathcal H}_{ij}(\theta),
\]
with $\bar{\mathcal H}_{ij}(\theta):=\frac{1}{2}\{\mathcal H_{ij}(\theta)+\mathcal H_{ji}(\theta)\}$,
hence $\mathcal H_N(\theta)\xrightarrow{p}E[\bar{\mathcal H}_{12}(\theta)] =
\Bigl(E[\mathcal H_{12}(\theta)]+E[\mathcal H_{21}(\theta)]\Bigr)/2
=
E[\mathcal H_{12}(\theta)]$ because \((X_1,X_2)\) and \((X_2,X_1)\) have the same distribution.
Combining pointwise convergence with stochastic equicontinuity and compactness of \(\Theta\), $\sup_{\theta\in\Theta}\|\mathcal H_N(\theta)-E\mathcal H_{12}(\theta)\|\xrightarrow{p}0.$
\end{proof}

\begin{theorem}\label{thm:pcle_an}
Under Assumptions~\ref{ass:parametric_compact}--\ref{ass:pcle_an},
$$\sqrt{N}\,(\widehat\theta-\theta_0)\xrightarrow{d}
N\!\left(0,\; \mathcal H_0^{-1}\Omega_0 \mathcal H_0^{-1}\right),$$
where
$\mathcal H_0
=
-\,E\!\left[e_{12}(\theta_0)\{1-e_{12}(\theta_0)\}R_{12}R_{12}'\right]$
and $\Omega_0:=4\,\var{E[\bar s_{12}| \Xi_1]},$ where $\Xi_i:=(X_i,U_i),$ and $\bar s_{12}:=(s_{12}(\theta_0)+s_{21}(\theta_0))/2.$
\end{theorem}

\begin{proof}
Define the score to be $s_N(\theta):=\partial_\theta L_N(\theta)
=
\frac{1}{N(N-1)}\sum_{i\ne j}s_{ij}(\theta).$
By Proposition~\ref{prop:pcle_consistency}, $\widehat\theta\xrightarrow{p}\theta_0$.
Choose $\varepsilon>0$ so that the ball $B(\theta_0,\varepsilon)$ lies in
$\operatorname{int}(\Theta)$ and in the differentiability neighborhood of
Assumption~\ref{ass:parametric_population}.
On the event $\|\widehat\theta-\theta_0\|<\varepsilon$, whose probability tends to one, the first-order
condition $s_N(\widehat\theta)=0$ holds and the entire parameter segment lies in
this ball. The fundamental theorem of calculus, applied componentwise, gives
\[
0=s_N(\theta_0)+\overline{\mathcal H}_N(\widehat\theta-\theta_0),
\qquad
\overline{\mathcal H}_N:=\int_0^1\mathcal H_N\bigl(\theta_0+t(\widehat\theta-\theta_0)\bigr)\,dt.
\]

Let $\mathcal H(\theta):=E[\mathcal H_{12}(\theta)]$. The logistic bound
$\|\mathcal H_{12}(\theta)\|\leq\|R_{12}\|^2/4$ and
Assumption~\ref{ass:parametric_moment} justify differentiation under the
expectation and continuity of $\mathcal H$ near $\theta_0$, with $\mathcal H(\theta_0)=\mathcal H_0$.
On the event above,
\begin{equation*}
\|\overline{\mathcal H}_N-\mathcal H_0\|
\leq\sup_{\theta\in\Theta}\|\mathcal H_N(\theta)-\mathcal H(\theta)\|+\sup_{\|\theta-\theta_0\|\leq\|\widehat\theta-\theta_0\|}
\|\mathcal H(\theta)-\mathcal H(\theta_0)\|.
\end{equation*}
Lemma~\ref{lem:hessian_unif}, consistency, and continuity imply
$\overline{\mathcal H}_N\xrightarrow{p}\mathcal H_0$. Since $\mathcal H_0$ is nonsingular,
$\overline{\mathcal H}_N^{-1}\xrightarrow{p}\mathcal H_0^{-1}$, and, with probability tending
to one,
$
\sqrt N\,(\widehat\theta-\theta_0)
=-\overline{\mathcal H}_N^{-1}\sqrt N\,s_N(\theta_0).
$

It remains to show that $\sqrt N\,s_N(\theta_0)\xrightarrow{d}N(0,\Omega_0)$. To this end, rewrite the score as a sum over unordered dyads:
\[
s_N(\theta_0)
=
\frac{2}{N(N-1)}\sum_{1\le i<j\le N}\bar s_{ij},
\]
with $\bar s_{ij}:=\frac{1}{2}\{s_{ij}(\theta_0)+s_{ji}(\theta_0)\}.$
By Assumption \ref{ass:pcle_an},
$E[s_{ij}(\theta_0)]
=
E\!\left[R_{ij}\{D_{ij}-E[D_{ij}| X_i,X_j]\}\right]
=0,$
and hence $E[\bar s_{12}]=0.$
Define $m_{ij}:=E[\bar s_{ij}| \Xi_i,\Xi_j]$ and $\psi_i:=E[m_{ij}| \Xi_i]$.
Then the score admits the decomposition
\[
s_N(\theta_0)=V_{1N}+V_{2N}+V_{3N},
\]
where
\[
V_{1N}:=\frac{2}{N}\sum_{i=1}^N\psi_i,
\]
\[
V_{2N}:=\frac{2}{N(N-1)}\sum_{1\le i<j\le N}\bigl(m_{ij}-\psi_i-\psi_j\bigr),
\]
and
\[
V_{3N}:=\frac{2}{N(N-1)}\sum_{1\le i<j\le N}\bigl(\bar s_{ij}-m_{ij}\bigr).
\]

I first study \(V_{1N}\). Since \(\{\Xi_i\}_{i=1}^N\) are i.i.d., \(\{\psi_i\}_{i=1}^N\) are i.i.d. as well.
Moreover, since \(R_{12}=R_{21}\), \(\bar s_{12}\) is bounded in norm by a constant times \(\|R_{12}\|\), so Jensen's inequality and
Assumption~\ref{ass:parametric_moment} imply $E\|\psi_1\|^{2+\delta}<\infty.$
Because \(E[\psi_1]=E[\bar s_{12}]=0\), the multivariate i.i.d.\ central limit theorem gives
\[
\sqrt N\,V_{1N}
=
2\cdot\frac{1}{\sqrt N}\sum_{i=1}^N\psi_i
\xrightarrow{d}
N(0,\Omega_0),
\qquad
\Omega_0=4\,\var{\psi_1}.
\]

Next consider \(V_{2N}\). Let $g_{ij}:=m_{ij}-\psi_i-\psi_j.$
Then $E[g_{ij}| \Xi_i]=0,$ and $E[g_{ij}| \Xi_j]=0.$
Fix any vector \(c\in\mathbb R^{\dim(\theta)}\). Since \(c'V_{2N}\) is scalar,
\[
E\bigl[(c'V_{2N})^2\bigr]
=
\frac{4}{N^2(N-1)^2}
\sum_{1\le i<j\le N}\sum_{1\le k<\ell\le N}
E\bigl[(c'g_{ij})(c'g_{k\ell})\bigr].
\]
If \(\{i,j\}\cap\{k,\ell\}=\varnothing\), then independence of the underlying \(\Xi\)'s and mean zero imply $E\bigl[(c'g_{ij})(c'g_{k\ell})\bigr]=0.$
If the dyads share exactly one index, say \(i=k\) and \(j\ne \ell\), then
$E\bigl[(c'g_{ij})(c'g_{i\ell})\bigr]
=
E\!\left[E[c'g_{ij}| \Xi_i]\;E[c'g_{i\ell}| \Xi_i]\right]
=0.$
Therefore only the diagonal terms remain, so
\[
E\bigl[(c'V_{2N})^2\bigr]
=
\frac{4}{N^2(N-1)^2}\sum_{1\le i<j\le N}E\bigl[(c'g_{ij})^2\bigr]
=
O(N^{-2}).
\]
Hence $\sqrt N\,c'V_{2N}\xrightarrow{p}0$
for every \(c\), and therefore $\sqrt N\,V_{2N}\xrightarrow{p}0.$

Finally consider \(V_{3N}\). Let $v_{ij}:=\bar s_{ij}-m_{ij}.$
Then $E[v_{ij}| \Xi_i,\Xi_j]=0.$
Conditional on \(\{\Xi_i\}_{i=1}^N\), the unordered dyads \(\{D_{ij}:1\le i<j\le N\}\) are mutually independent under the CID model. Since \(v_{ij}\) depends on the sample only through
\((D_{ij},\Xi_i,\Xi_j)\), it follows that conditional on \(\{\Xi_i\}_{i=1}^N\), the collection \(\{v_{ij}:1\le i<j\le N\}\) is mutually independent with mean zero. Thus, for any fixed \(c\),
\[
E\bigl[(c'V_{3N})^2| \Xi_1,\dots,\Xi_N\bigr]
=
\frac{4}{N^2(N-1)^2}\sum_{1\le i<j\le N}E\bigl[(c'v_{ij})^2| \Xi_1,\dots,\Xi_N\bigr].
\]
Taking expectations, $E\bigl[(c'V_{3N})^2\bigr]
=
O(N^{-2}).$
Hence
$\sqrt N\,c'V_{3N}\xrightarrow{p}0$
for every \(c\), and therefore
$\sqrt N\,V_{3N}\xrightarrow{p}0.$

Combining the three pieces,
$\sqrt N\,s_N(\theta_0)
=
\sqrt N\,V_{1N}+\sqrt N\,V_{2N}+\sqrt N\,V_{3N}
\xrightarrow{d}
N(0,\Omega_0).$
Substituting this limit into the integral-Hessian expansion and applying Slutsky's theorem gives $\sqrt N(\widehat\theta-\theta_0)
\xrightarrow{d}
N\!\left(0,\;\mathcal H_0^{-1}\Omega_0 \mathcal H_0^{-1}\right)$ as desired.
\end{proof}

\subsection{Nonparametric estimation of the subgraph moments}\label{subsec:nonparametric-moments}
    With continuously distributed covariates, the empirical density $P_N(S,x)$ in \eqref{eq:defP_N} is uninformative, because exact covariate matches occur with probability zero. I therefore smooth over nearby covariate values. Recall that $d_W=pd_X$ is the dimension of the stacked covariate vector of a $p$-tuple, $K:\mathbb{R}^{d_W}\to\mathbb{R}$ is a fixed multivariate kernel, and $h_N>0$ is a bandwidth sequence with $h_N\to 0$. The kernel estimator is
    \[
\widehat P_N(S,x)
=
\frac{
\sum_{\mathbf i_p\in\mathcal P_{p,N}}
K\!\left(\frac{x-X_{\mathbf i_p}}{h_N}\right)
\mathbf 1\{G_N[\mathbf i_p]\simeq S\}
}{
\sum_{\mathbf i_p\in\mathcal P_{p,N}}
K\!\left(\frac{x-X_{\mathbf i_p}}{h_N}\right)
}.
\]

This estimator is the nonparametric fit of the network moment \(P(S,x)\).
It is the subgraph analogue of the Nadaraya--Watson estimator. \textcite{ding_nonparametric_2026} establishes a variance bound, uniform convergence rates, and pointwise asymptotic normality for it; I use the uniform rates here.

Recall that \(Y_{\mathbf i_p,S}=\mathbf 1\{G_N[\mathbf i_p]\simeq S\}\) is the response of the tuple \(\mathbf i_p\) and \(X_{\mathbf i_p}\in\mathbb R^{d_W}\) is its stacked covariate vector.
The regression function of \(Y_{\mathbf i_p,S}\) is \(m_S(w):=E[Y_{\mathbf i_p,S}\mid X_{\mathbf i_p}=w]=P(S,w)\).
Recall that \(\mathcal P_{p,N}\) contains the ordered tuples of distinct
vertices, \(M_N:=|\mathcal P_{p,N}|=N!/(N-p)!\), and
\(\mathcal C_{p,N}\) contains their unordered vertex sets; for
\(\mathbf q_p\in\mathcal C_{p,N}\), \(\mathfrak S(\mathbf q_p)\) denotes its \(p!\) permutations.
As in Theorem~\ref{thm:pcle_an}, \(\Xi_i:=(X_i,U_i)\) denotes the i.i.d.\ vertex variables of the CID
model; write \(\Xi_{\mathbf q_p}:=(\Xi_q:q\in\mathbf q_p)\).
Let \(f\) be the marginal density of \(X_i\); the stacked covariates then have density
\[
f_W(w)=\prod_{a=1}^p f(w_a),
\qquad w=(w_1',\ldots,w_p')',\quad w_a\in\mathbb R^{d_X}.
\]

Recall that \(K_h(v)=h_N^{-d_W}K(v/h_N)\), as defined after \eqref{eq:test_stats}, and set
\[
K_{\mathbf i_p}(w):=K_h(w-X_{\mathbf i_p}).
\]
The normalized numerator and common density estimator are
\[
\widehat\Psi_N(w):=\frac1{M_N}
\sum_{\mathbf i_p\in\mathcal P_{p,N}}
K_{\mathbf i_p}(w)Y_{\mathbf i_p,S},
\qquad
\widehat f_N(w):=\frac1{M_N}
\sum_{\mathbf i_p\in\mathcal P_{p,N}}K_{\mathbf i_p}(w).
\]
They estimate \(\Psi(w):=P(S,w)f_W(w)\) and \(f_W(w)\), respectively,
and \(\widehat P_N(S,w)=\widehat\Psi_N(w)/\widehat f_N(w)\).
In the regression notation of \textcite{ding_nonparametric_2026}, this is
\(\widehat m_N(w)\) with response \(Y_{\mathbf i_p}=Y_{\mathbf i_p,S}\) and
regression function \(m(w)=P(S,w)\). That paper writes the kernel weight as
\(h_N^{-d_W}K((X_{\mathbf i_p}-w)/h_N)\); its results apply here with
\(K(-\,\cdot)\) in place of \(K\), which satisfies the same assumptions.

\begin{assumption}[Covariate density]\label{aspt:Model}
The covariates have a marginal density \(f\) satisfying
\(\sup_{x\in\mathbb R^{d_X}}f(x)\le B_1<\infty\) (without loss of generality, \(B_1\ge1\)).
\end{assumption}
The CID model fits the framework of \textcite[Section~2]{ding_nonparametric_2026}:
writing \(D_{ij}=\mathbf 1\{V_{ij}\le h(X_i,X_j,U_i,U_j)\}\) with i.i.d.\
uniform shocks \(V_{ij}\) independent of the vertex variables,
\(Y_{\mathbf i_p,S}\) is a fixed function of \(\Xi_{\mathbf i_p}\) and the
shocks of the dyads in \(\mathbf i_p\), as in its Example~2.1. In particular,
subgraphs with disjoint vertex sets are independent.
Since \(0\le Y_{\mathbf i_p,S}\le1\), the response has bounded conditional
moments of every order, so the moment conditions in its Assumptions~3.1(ii)
and~3.3(i) hold automatically.\footnote{The density-weighted conditional cross
moments are bounded by \(B_1^{2p-1}\), and the conditional \(s\)-th moments
by \(B_1^p\) for every \(s>2\).}

\begin{assumption}[Kernel]\label{aspt:KernelA}
The kernel \(K:\mathbb R^{d_W}\to\mathbb R\) satisfies:
\begin{enumerate}[label=(\roman*)]
\item \label{asptpart:KernelA_1}
\(\int_{\mathbb R^{d_W}}|K(u)|\,du\le C_1<\infty\).
\item \label{asptpart:KernelA_2}
For every \(A\subseteq\{1,\ldots,p\}\),
$
\sup_{u_A}\int |K(u_1,\ldots,u_p)|\,du_{A^c}\le C_2<\infty,
$
where \(u_a\in\mathbb R^{d_X}\), \(u_A=(u_a:a\in A)\), and
\(du_{A^c}\) integrates the remaining blocks.
\item \label{asptpart:KernelA_3}
\(\sup_{u\in\mathbb R^{d_W}}|K(u)|\le K_{\max}<\infty\).
\end{enumerate}
\end{assumption}

\begin{assumption}[Kernel regularity]\label{aspt:regularity}
For finite constants \(L_K,\Lambda_1\), the kernel satisfies
\(K(u)=0\) when \(\|u\|_\infty>L_K\) and
$
|K(u)-K(v)|\le\Lambda_1\|u-v\|_\infty
$ for $u,v\in\mathbb R^{d_W}.$
\end{assumption}

\begin{definition}[H\"older smoothness]\label{aspt:holder}
Let \(\beta,L>0\), let \(\lfloor\beta\rfloor:=\max\{r\in\mathbb Z_{\ge0}:r<\beta\}\)
be the greatest integer strictly less than \(\beta\), and let
\(\alpha:=\beta-\lfloor\beta\rfloor\in(0,1]\).
For a multi-index \(s=(s_1,\ldots,s_{d_W})\), let
\(|s|:=\sum_j s_j\) and
\(D^s:=\partial^{|s|}/(\partial w_1^{s_1}\cdots\partial w_{d_W}^{s_{d_W}})\).
A function \(a:\mathbb R^{d_W}\to\mathbb R\) belongs to
\(\Sigma(\beta,L)\) if it is \(\lfloor\beta\rfloor\) times continuously differentiable and
\[
|D^s a(w)-D^s a(w')|\le L\|w-w'\|_\infty^\alpha
\]
for every \(|s|=\lfloor\beta\rfloor\) and every \(w,w'\in\mathbb R^{d_W}\).
\end{definition}

\begin{assumption}[Higher-order kernel]\label{aspt:kernelB}
The kernel satisfies \(\int_{\mathbb R^{d_W}}K(u)\,du=1\) and
\[
\int_{\mathbb R^{d_W}}u^rK(u)\,du=0
\quad\text{for every multi-index }r\in\mathbb Z_+^{d_W}
\text{ with }1\le |r|<\beta,
\]
where \(u^r:=u_1^{r_1}\cdots u_{d_W}^{r_{d_W}}\).
Also, \(\int_{\mathbb R^{d_W}}\|u\|_\infty^\beta|K(u)|\,du<\infty\).
\end{assumption}
Assumption~\ref{aspt:regularity} is \textcite[Assumption~3.3(ii)]{ding_nonparametric_2026}.
Assumptions~\ref{aspt:Model}, \ref{aspt:KernelA}, and~\ref{aspt:kernelB}
are its Assumptions~3.1(i), 3.2, and~3.4, and Definition~\ref{aspt:holder}
is its Definition~2.1.

For later use, define the stochastic and bias-adjusted rates
\begin{equation}\label{eq:np-rates}
a_N:=\left(\frac{\ln N}{Nh_N^{d_X}}\right)^{1/2},
\qquad a_N^*:=a_N+h_N^\beta.
\end{equation}
Under the regularity and bandwidth conditions maintained in
Assumption~\ref{ass:HM-complete}, the uniform convergence results in
\textcite[Theorems~3.2 and~3.3]{ding_nonparametric_2026} give
\begin{equation}\label{eq:np-uniform-rates}
\sup_{w\in\mathcal W_\pi}|\widehat f_N(w)-f_W(w)|=O_p(a_N^*),
\qquad
\sup_{w\in\mathcal W_\pi}|\widehat P_N(S,w)-P(S,w)|=O_p(a_N^*),
\end{equation}
where \(\mathcal W_\pi\) is the compact support of the test weight and
\(f_W\) is bounded away from zero there. The density rate is Theorem~3.2
applied with \(Y_{\mathbf i_p}\equiv1\), combined with the smoothing-bias
bound in its Lemma~E.1, as in the proof of its Theorem~3.3. The rate for
\(\widehat P_N\) is its Theorem~3.3 with a fixed ball containing
\(\mathcal W_\pi\); the proof of that theorem uses the lower bound on
\(f_W\) only over the set on which the supremum is taken, so the bound on
\(\mathcal W_\pi\) suffices.

\section{Asymptotics of the test statistic}
\label{sec:test-stat-asymptotics}

In this section I write \(w\in\mathbb R^{d_W}\) for the stacked covariate argument, written \(x\) in Section~\ref{sec:test}. Recall that the test statistic is
\[
        T_{N,S}
        :=
        Nh_N^{d_X/2}
        \int
        \left\{
        \widehat P_N(S,w)
        -
        \mathcal K_{h,N}P(S,\cdot;e_{\widehat\theta})(w)
        \right\}^2
        \pi(w)\,dw.
\]

Define the subgraph probability conditional on the vertex variables, which averages only over the dyad shocks $V_{ij}$:
\[
        \mu_S(\xi_1,\ldots,\xi_p)
        :=
        E\left[
        Y_{(1,\ldots,p),S}
        \mid
        \Xi_1=\xi_1,\ldots,\Xi_p=\xi_p
        \right].
\]
Thus
        $P(S,x_1,\ldots,x_p)
        =
        E\left[
        \mu_S(\Xi_1,\ldots,\Xi_p)
        \mid
        X_1=x_1,\ldots,X_p=x_p
        \right].$

To define the first-order Hoeffding projection, fix a value \(\xi=(x,u)\) of a vertex variable. For \(a=1,\ldots,p\), let
\(\Xi^{(a)}(\xi):=(\Xi_1,\ldots,\Xi_{a-1},\xi,\Xi_{a+1},\ldots,\Xi_p)\)
and
\(X^{(a)}(x):=(X_1,\ldots,X_{a-1},x,X_{a+1},\ldots,X_p)\). Define
\begin{equation}
\label{eq:ell-explicit}
        \ell_{N,S}(\xi;w)
        :=
        \frac{1}{p f_W(w)}
        \sum_{a=1}^p
        E\left[
        K_h\{w-X^{(a)}(x)\}                                        \times
        \left\{
        \mu_S\bigl(\Xi^{(a)}(\xi)\bigr)
        -
        P\bigl(S,X^{(a)}(x)\bigr)
        \right\} \right],
\end{equation}
where the expectation integrates over \(\{\Xi_b:b\neq a\}\). By construction, $E[\ell_{N,S}(\Xi_1;w)]=0.$

Define
\[
        \Gamma_{N,S}(w,z)
        :=
        p^2h_N^{d_X}
        E\left[
        \ell_{N,S}(\Xi_1;w)
        \ell_{N,S}(\Xi_1;z)
        \right],
\]
\[
        b_{N,S}
        :=
        h_N^{-d_X/2}
        \int
        \Gamma_{N,S}(w,w)\pi(w)\,dw,
\]
and
\[
        \mathcal V_{N,S}
        :=
        2h_N^{-d_X}
        \int\!\!\int
        \Gamma_{N,S}(w,z)^2
        \pi(w)\pi(z)\,dw\,dz.
\]

\begin{assumption}[Regularity conditions]
\label{ass:HM-complete}
The CID model and Assumptions~\ref{ass:parametric_compact}--\ref{ass:pcle_an}, \ref{aspt:Model}, \ref{aspt:KernelA}, \ref{aspt:regularity}, and \ref{aspt:kernelB} hold. In addition, the following conditions hold.

\begin{enumerate}[label=(\roman*)]

\item
The weight \(\pi\) is nonnegative and bounded with compact support \(\mathcal W_\pi\), and there is a constant \(\underline f>0\) such that \(\inf_{w\in\mathcal W_\pi}f_W(w)\geq\underline f\).

\item
 The functions \(f_W\), \(P(S,\cdot)\), and the first two derivatives of \(P(S,\cdot;e_\theta)\) with respect to \(\theta\) belong to \(\Sigma(\beta,L)\), uniformly for \(\theta\) in a neighborhood of \(\theta_0\).

\item
The bandwidth satisfies \(h_N\to0\) and \(a_N^*h_N^{-d_X/2}\to0\).
Equivalently, \(Nh_N^{2d_X}/\ln N\to\infty\) and \(\beta>d_X/2\).

\item
The variance constant \(\mathcal V_{N,S}\) converges to a limit \(\mathcal V_S\in(0,\infty)\).

\end{enumerate}
\end{assumption}

\begin{theorem}[Asymptotic null distribution]\label{thm:null-distribution}
Suppose Assumption~\ref{ass:HM-complete} holds. Then, under \(H_0\),
\[
        T_{N,S}-b_{N,S}
        \to_d
        N(0,\mathcal V_S).
\]
\end{theorem}

\begin{proof}
Throughout the proof, \(C\) denotes a finite constant that does not depend on
\(N\) and may change from line to line. Under \(H_0\), set $ \eta_{\mathbf i_p}
        :=
        Y_{\mathbf i_p,S}-P(S,X_{\mathbf i_p}),$ so that
\[
        Y_{\mathbf i_p,S}
        -
        P(S,X_{\mathbf i_p};e_{\widehat\theta})
        =
        \eta_{\mathbf i_p}
        -
        \left\{
        P(S,X_{\mathbf i_p};e_{\widehat\theta})
        -
        P(S,X_{\mathbf i_p};e_{\theta_0})
        \right\}.
\]

Define
\[
        U_{N1}(w)
        :=
        \frac{1}{M_Nf_W(w)}
        \sum_{\mathbf i_p\in\mathcal P_{p,N}}
        K_h(w-X_{\mathbf i_p})\eta_{\mathbf i_p},
\]
\[
        U_{N2}(w)
        :=
        -\frac{1}{M_Nf_W(w)}
        \sum_{\mathbf i_p\in\mathcal P_{p,N}}
        K_h(w-X_{\mathbf i_p})
        \left\{
        P(S,X_{\mathbf i_p};e_{\widehat\theta})
        -
        P(S,X_{\mathbf i_p};e_{\theta_0})
        \right\}.
\]
Since the nonparametric
estimator and the smoothed parametric estimator have the same denominator,
\begin{equation}
\label{eq:exact-ratio}
        \widehat P_N(S,w)
        -
        \mathcal K_{h,N}P(S,\cdot;e_{\widehat\theta})(w)
        =
        \frac{f_W(w)}{\widehat f_N(w)}
        \{U_{N1}(w)+U_{N2}(w)\}.
\end{equation}

\medskip
\noindent
\textit{Step 1: Replacement of the random denominator.}

Let
\(\delta_N:=\|\widehat f_N-f_W\|_{\infty,\mathcal W_\pi}\). By \eqref{eq:np-uniform-rates},
\(\delta_N=O_p(a_N^*)\) and
\(\delta_Nh_N^{-d_X/2}=o_p(1)\). Since \(f_W\geq\underline f\) on
\(\mathcal W_\pi\),
\[
        \Pr\left(
        \inf_{w\in\mathcal W_\pi}\widehat f_N(w)
        \geq\frac{\underline f}{2}
        \right)\to1.
\]
On this event, equation \eqref{eq:exact-ratio} becomes
\[
        \widehat P_N(S,w)
        -
        \mathcal K_{h,N}P(S,\cdot;e_{\widehat\theta})(w)
        =
        U_{N1}(w)+U_{N2}(w)+R_{f,N}(w),
\]
where
\[
        R_{f,N}(w)
        =
        \{U_{N1}(w)+U_{N2}(w)\}
        \left\{
        \frac{f_W(w)}{\widehat f_N(w)}-1
        \right\}.
\]
Moreover,
\[
        \left|\frac{f_W(w)}{\widehat f_N(w)}-1\right|
        =\frac{|f_W(w)-\widehat f_N(w)|}{\widehat f_N(w)}
        \leq\frac{2}{\underline f}\delta_N.
\]
Thus,
\begin{equation}
\label{eq:Rf-envelope}
        |R_{f,N}(w)|
        \leq
        \frac{2}{\underline f}\,\delta_N
        \{|U_{N1}(w)|+|U_{N2}(w)|\}.
\end{equation}

I first bound the integrated square of \(U_{N1}\). Because
\(E[\eta_{\mathbf i_p}\mid X_{\mathbf i_p}]=0\), disjoint tuples have zero
covariance. There are \(O(N^{2p-t})\) ordered pairs of tuples sharing exactly
\(t\) vertices. For such a pair, integration over the \(2p-t\) distinct
vertex covariates, as in the proof of
\textcite[Theorem~3.1]{ding_nonparametric_2026}, yields
\[
        E\int
        \left|
        K_h(w-X_{\mathbf i_p})
        K_h(w-X_{\mathbf k_p})
        \eta_{\mathbf i_p}\eta_{\mathbf k_p}
        \right|
        \pi(w)\,dw
        \leq
        Ch_N^{-td_X}.
\]
Since \(M_N^2\asymp N^{2p}\),
\begin{equation*}
        E\int U_{N1}(w)^2\pi(w)\,dw
        \leq
        C\sum_{t=1}^p
        N^{-t}h_N^{-td_X} =
        O\left(\frac1{Nh_N^{d_X}}\right),
\end{equation*}
where the \(t=1\) term dominates because \(Nh_N^{d_X}\to\infty\).
Consequently,
\begin{equation}
\label{eq:scaled-Un1-order}
        Nh_N^{d_X/2}
        \int U_{N1}(w)^2\pi(w)\,dw
        =
        O_p(h_N^{-d_X/2}).
\end{equation}

Step 2 below shows that
\begin{equation}
\label{eq:Un2-square}
        Nh_N^{d_X/2}
        \int U_{N2}(w)^2\pi(w)\,dw=o_p(1).
\end{equation}
Using \eqref{eq:Rf-envelope} and
\((a+b)^2\leq2a^2+2b^2\),
\[
\begin{aligned}
        Nh_N^{d_X/2}
        \int R_{f,N}(w)^2\pi(w)\,dw
        &\leq
        C\delta_N^2
        Nh_N^{d_X/2}
        \int
        \{U_{N1}(w)^2+U_{N2}(w)^2\}
        \pi(w)\,dw                                      \\
        &=
        O_p(\delta_N^2h_N^{-d_X/2})
        =
        o_p(1).
\end{aligned}
\]
The last conclusion follows because
\(\delta_Nh_N^{-d_X/2}=o_p(1)\) and \(\delta_N=o_p(1)\).

Similarly, \eqref{eq:Rf-envelope}, \eqref{eq:scaled-Un1-order}, and \eqref{eq:Un2-square} imply
\[
\begin{aligned}
        &Nh_N^{d_X/2}
        \int
        |R_{f,N}(w)|
        \{|U_{N1}(w)|+|U_{N2}(w)|\}
        \pi(w)\,dw                                      \\
        &\qquad\leq
        C\delta_N
        Nh_N^{d_X/2}
        \int
        \{|U_{N1}(w)|+|U_{N2}(w)|\}^2
        \pi(w)\,dw                                      \\
        &\qquad\leq
        2C\delta_N
        Nh_N^{d_X/2}
        \int
        \{U_{N1}(w)^2+U_{N2}(w)^2\}
        \pi(w)\,dw                                      \\
        &\qquad=
        O_p(\delta_Nh_N^{-d_X/2})
        =
        o_p(1).
\end{aligned}
\]
Therefore,
\begin{equation}
\label{eq:HM-after-density}
        T_{N,S}
        =
        Nh_N^{d_X/2}
        \int
        \{U_{N1}(w)+U_{N2}(w)\}^2
        \pi(w)\,dw
        +
        o_p(1).
\end{equation}

\medskip
\noindent
\textit{Step 2: The squared parametric estimation deviation.}

Only tuples with \(X_{\mathbf i_p}\) within \(L_Kh_N\) of \(\mathcal W_\pi\)
contribute to \(U_{N2}(w)\) for \(w\in\mathcal W_\pi\), because \(K\) vanishes
outside \(\{\|u\|_\infty\le L_K\}\) by Assumption~\ref{aspt:regularity} and
\(\pi\) is supported on the compact set \(\mathcal W_\pi\). On this compact
neighborhood of \(\mathcal W_\pi\), the first two \(\theta\)-derivatives of
\(P(S,\cdot\,;e_\theta)\), which are continuous by
Assumption~\ref{ass:HM-complete}(ii), are bounded uniformly for \(\theta\) in a
neighborhood of \(\theta_0\).
A second-order Taylor expansion in \(\theta\) gives
\[
        P(S,X_{\mathbf i_p};e_{\widehat\theta})
        -
        P(S,X_{\mathbf i_p};e_{\theta_0})
        =
        \partial_\theta P(S,X_{\mathbf i_p};e_{\theta_0})'
        (\widehat\theta-\theta_0)
        +
        R_{\mathbf i_p}^{\theta},
\]
where, uniformly over \(\mathbf i_p\),
\[
        |R_{\mathbf i_p}^{\theta}|
        \leq
        C\|\widehat\theta-\theta_0\|^2
        =
        O_p(N^{-1}).
\]

Let
\[
        A_N(w)
        :=
        \frac1{M_N}
        \sum_{\mathbf i_p\in\mathcal P_{p,N}}
        \frac{
        K_h(w-X_{\mathbf i_p})
        \partial_\theta P(S,X_{\mathbf i_p};e_{\theta_0})
        }{
        f_W(w)
        },
\]
and define \(R_N^\theta(w)\) analogously with
\(R_{\mathbf i_p}^{\theta}\) in place of
\(\partial_\theta P(S,X_{\mathbf i_p};e_{\theta_0})\). Then
\begin{equation}
\label{eq:Un2-expansion}
        U_{N2}(w)
        =
        -A_N(w)'(\widehat\theta-\theta_0)
        -
        R_N^\theta(w).
\end{equation}

Write $A_N(w)
        =
        \overline A_N(w)+\widetilde A_N(w)$ with $\overline A_N(w):=E[A_N(w)]$.
Boundedness of \(\partial_\theta P(S,\cdot\,;e_{\theta_0})\), \(f_W^{-1}\), and
\(\int|K_h(w-z)|\,dz\) implies $\sup_{w\in\mathcal W_\pi}
        |\overline A_N(w)|
        \leq C.$
I next calculate the integrated variance of \(\widetilde A_N\). Expanding the
square gives a double sum over ordered tuples. Terms indexed by disjoint tuples
have zero covariance because they depend on disjoint sets of i.i.d. vertex
covariates. If two tuples share \(t\geq1\) vertices, partial kernel integration
gives an integrated covariance of order at most \(h_N^{-td_X}\). Hence
\begin{equation*}
        E\|\widetilde A_N\|_{L^2(\pi)}^2
        \leq
        \frac{C}{M_N^2}
        \sum_{t=1}^p
        N^{2p-t}h_N^{-td_X}
        \leq
        C\sum_{t=1}^p
        N^{-t}h_N^{-td_X}
        =
        O\left(\frac1{Nh_N^{d_X}}\right).
\end{equation*}

The final equality follows because the \(t=1\) term dominates.
Thus $\|\widetilde A_N\|_{L^2(\pi)}
        =
        O_p\left((Nh_N^{d_X})^{-1/2}\right)=o_p(1).$
Since \(\pi\) is bounded with compact support,
\(\|\overline A_N\|_{L^2(\pi)}\le C\), and hence
\(\|A_N\|_{L^2(\pi)}=O_p(1)\).

Also, the average absolute kernel weight \(M_N^{-1}\sum_{\mathbf i_p}|K_h(w-X_{\mathbf i_p})|\) is \(O_p(1)\) uniformly over \(\mathcal W_\pi\) by \eqref{eq:np-uniform-rates} applied with \(|K|\) in place of \(K\), which together with \(\sup_{\mathbf i_p}|R_{\mathbf i_p}^\theta|=O_p(N^{-1})\) implies
$ \sup_{w\in\mathcal W_\pi}|R_N^\theta(w)|
        =
        O_p(N^{-1}).$
Using \eqref{eq:Un2-expansion}, \((a+b)^2\le2a^2+2b^2\), and
\(\|\widehat\theta-\theta_0\|=O_p(N^{-1/2})\) from Theorem~\ref{thm:pcle_an},
\[
        \int U_{N2}(w)^2\pi(w)\,dw
        \le
        2\|\widehat\theta-\theta_0\|^2\|A_N\|_{L^2(\pi)}^2
        +
        2\|R_N^\theta\|_{L^2(\pi)}^2
        =
        O_p(N^{-1}),
\]
and therefore, establishing \eqref{eq:Un2-square},
        $Nh_N^{d_X/2}
        \int U_{N2}(w)^2\pi(w)\,dw
        =
        O_p(h_N^{d_X/2})
        =
        o_p(1).$

\medskip
\noindent
\textit{Step 3: The nonparametric--parametric cross term.}

By \eqref{eq:Un2-expansion}, the nonparametric--parametric cross term is
\[
        \int U_{N1}(w)U_{N2}(w)\pi(w)\,dw
        =
        -(\widehat\theta-\theta_0)'
        \int U_{N1}(w)A_N(w)\pi(w)\,dw
        -
        \int U_{N1}(w)R_N^\theta(w)\pi(w)\,dw.
\]
I bound the first term again using the decomposition
\(A_N=\overline A_N+\widetilde A_N\).
Fix a coordinate \(m\in\{1,\ldots,\dim(\theta)\}\). Substituting the definition of \(U_{N1}\),
\[
\begin{aligned}
        \int
        U_{N1}(w)\overline A_{N,m}(w)
        \pi(w)\,dw
        &=
        \frac1{M_N}
        \sum_{\mathbf i_p\in\mathcal P_{p,N}}
        \eta_{\mathbf i_p}
        \int
        \frac{
        K_h(w-X_{\mathbf i_p})
        \overline A_{N,m}(w)\pi(w)
        }{
        f_W(w)
        }
        \,dw.
\end{aligned}
\]
The inner integral is uniformly bounded in \(X_{\mathbf i_p}\), because
\(\overline A_N\), \(\pi\), and \(f_W^{-1}\) are bounded and the
\(L^1\)-norm of \(K_h\) is constant in \(h_N\). Each summand is centered since
\(E[\eta_{\mathbf i_p}\mid X_{\mathbf i_p}]=0\).

Pairs of disjoint tuples have zero covariance. For each \(t\geq1\), there are
\(O(N^{2p-t})\) ordered pairs sharing exactly \(t\) vertices, and the covariance
of the corresponding bounded summands is bounded by a constant. Therefore,
using \(M_N^2\asymp N^{2p}\),
\begin{equation*}
        \var{\int U_{N1}(w)\overline A_{N,m}(w)\pi(w)\,dw}
        \leq
        \frac{C}{M_N^2}
        \sum_{t=1}^pN^{2p-t}
        \leq
        C N^{-2p}
        \sum_{t=1}^pN^{2p-t}
        =
        C\sum_{t=1}^pN^{-t}
        =
        O(N^{-1}).
\end{equation*}
Since this holds for every coordinate \(m\),
\[
        \int
        U_{N1}(w)\overline A_N(w)\pi(w)\,dw
        =
        O_p(N^{-1/2}).
\]

For the random part, Cauchy--Schwarz gives
\[
\begin{aligned}
        \left|
        \int
        U_{N1}(w)\widetilde A_N(w)\pi(w)\,dw
        \right|
        &\leq
        \|U_{N1}\|_{L^2(\pi)}
        \|\widetilde A_N\|_{L^2(\pi)}                    \\
        &=
        O_p\left((Nh_N^{d_X})^{-1/2}\right)
        O_p\left((Nh_N^{d_X})^{-1/2}\right)
       =
        O_p\left(\frac1{Nh_N^{d_X}}\right).
\end{aligned}
\]
The two norm orders follow respectively from
the integrated second-moment bound for \(U_{N1}\) in Step~1 and the integrated variance calculation for
\(\widetilde A_N\). Hence
\begin{equation}
\label{eq:smooth-functional}
        \int
        U_{N1}(w)A_N(w)\pi(w)\,dw
        =
        O_p(N^{-1/2})
        +
        O_p\left(\frac1{Nh_N^{d_X}}\right).
\end{equation}

Combining \eqref{eq:Un2-expansion} and
\eqref{eq:smooth-functional}, and using
\(\|\widehat\theta-\theta_0\|=O_p(N^{-1/2})\),
\[
\begin{aligned}
        &Nh_N^{d_X/2}
        \left|
        (\widehat\theta-\theta_0)'
        \int U_{N1}(w)A_N(w)\pi(w)\,dw
        \right|                                         \\
        &\qquad\leq
        Nh_N^{d_X/2}
        O_p(N^{-1/2})
        \left\{
        O_p(N^{-1/2})
        +
        O_p\left(\frac1{Nh_N^{d_X}}\right)
        \right\}                                        \\
        &\qquad=
        O_p(h_N^{d_X/2})
        +
        O_p\left(\frac1{\sqrt{Nh_N^{d_X}}}\right)
        =
        o_p(1).
\end{aligned}
\]
For the Taylor remainder,
\[
\begin{aligned}
        Nh_N^{d_X/2}
        \left|
        \int
        U_{N1}(w)R_N^\theta(w)\pi(w)\,dw
        \right|
        &\leq
        Nh_N^{d_X/2}
        \|U_{N1}\|_{L^2(\pi)}
        \|R_N^\theta\|_{L^2(\pi)}                        \\
        &=
        Nh_N^{d_X/2}
        O_p\left((Nh_N^{d_X})^{-1/2}\right)
        O_p(N^{-1})                                     \\
        &=
        O_p(N^{-1/2})
        =
        o_p(1).
\end{aligned}
\]
Consequently, $  Nh_N^{d_X/2}
        \int
        U_{N1}(w)U_{N2}(w)\pi(w)\,dw
        =
        o_p(1).$
Together with \eqref{eq:HM-after-density} and \eqref{eq:Un2-square}, this gives
\begin{equation}
\label{eq:reduced-to-Un1}
        T_{N,S}
        =
        Nh_N^{d_X/2}
        \int U_{N1}(w)^2\pi(w)\,dw
        +
        o_p(1).
\end{equation}

\medskip
\noindent
\textit{Step 4: Reduction of $U_{N1}$ to its H\'ajek projection.}

 For
\(\mathbf q_p\in\mathcal C_{p,N}\), define the symmetrized residualized kernel
\[
        \Upsilon^0_{N,\mathbf q_p}(w)
        :=
        \frac1{p!}
        \sum_{\sigma\in\mathfrak S(\mathbf q_p)}
        \frac{
        K_h(w-X_{\sigma(\mathbf q_p)})
        }{
        f_W(w)
        }
        \times
        \left\{
        Y_{\sigma(\mathbf q_p),S}
        -
        P(S,X_{\sigma(\mathbf q_p)})
        \right\}.
\]
Then $U_{N1}(w)=\binom Np^{-1}\sum_{\mathbf q_p\in\mathcal C_{p,N}}\Upsilon^0_{N,\mathbf q_p}(w)$.
To average out the dyad shocks, set
\(\overline \Upsilon^0_{N,\mathbf q_p}(w)
:=E[\Upsilon^0_{N,\mathbf q_p}(w)\mid\Xi_{\mathbf q_p}]\). By the definition of
\(\mu_S\),
\[
        \overline \Upsilon^0_{N,\mathbf q_p}(w)
        =
        \frac1{p!}
        \sum_{\sigma\in\mathfrak S(\mathbf q_p)}
        \frac{
        K_h(w-X_{\sigma(\mathbf q_p)})
        }{
        f_W(w)
        }
\times
        \left\{
        \mu_S(\Xi_{\sigma(\mathbf q_p)})
        -
        P(S,X_{\sigma(\mathbf q_p)})
        \right\}.
\]
Moreover, \(E[\overline \Upsilon^0_{N,\mathbf q_p}(w)]=0\). Write
\[
        U_{N1}(w)
        =
        \binom Np^{-1}
        \sum_{\mathbf q_p\in\mathcal C_{p,N}}
        \overline \Upsilon^0_{N,\mathbf q_p}(w)
        +
        V_{N,S}^{e}(w),
\]
where \(V_{N,S}^{e}\), the edge-shock remainder, is the average of
\(\Upsilon^0_{N,\mathbf q_p}-\overline \Upsilon^0_{N,\mathbf q_p}\), so that $E[
        V_{N,S}^{e}(w)
        \mid
        \Xi_1,\ldots,\Xi_N
        ]
        =
        0.$
The first term is an order-\(p\) U-statistic in the i.i.d. vertex variables, and the Hoeffding decomposition of this U-statistic is $ \sum_{k=1}^pU_{N,S,k}(w)$ with
\[
        U_{N,S,k}(w)
        :=
        \frac{\binom pk}{\binom Nk}
        \sum_{\mathbf q_k\in\mathcal C_{k,N}}
        \phi^0_{N,S,k}(\Xi_{\mathbf q_k};w).
\]
Here
\(\phi^0_{N,S,k}\) is the canonical kernel of order \(k\), which I now define.
For \(A\subseteq\{1,\ldots,p\}\) and \(\xi_A:=(\xi_a)_{a\in A}\), let
\[
        \psi_{N,S,A}(\xi_A;w)
        :=
        E\left[
        \overline \Upsilon^0_{N,\{1,\ldots,p\}}(w)
        \mid
        \Xi_a=\xi_a,\ a\in A
        \right]
\]
be the projection of the kernel onto the vertices in \(A\). Since
\(\overline \Upsilon^0_{N,\{1,\ldots,p\}}(w)\) is a symmetric function of
\(\Xi_1,\ldots,\Xi_p\), \(\psi_{N,S,A}\) depends on \(A\) only through
\(|A|\), and \(\psi_{N,S,\varnothing}(w)=E[\overline \Upsilon^0_{N,\{1,\ldots,p\}}(w)]=0\).
The canonical kernels are obtained by M\"obius inversion of the
\(\psi_{N,S,A}\) over subsets,
\begin{equation}
\label{eq:Mobius-kernel}
        \phi^0_{N,S,k}(\xi_1,\ldots,\xi_k;w)
        :=
        \sum_{A\subseteq\{1,\ldots,k\}}
        (-1)^{k-|A|}
        \psi_{N,S,A}(\xi_A;w),
        \qquad k=1,\ldots,p.
\end{equation}
They are canonical in the sense that \(\phi^0_{N,S,k}\) has conditional mean
zero given any \(k-1\) of its arguments, as I verify below for \(k\geq2\).

For \(k=1\), \eqref{eq:Mobius-kernel} gives
\(\phi^0_{N,S,1}(\xi;w)=\psi_{N,S,\{1\}}(\xi;w)=E[\overline \Upsilon^0_{N,\{1,\ldots,p\}}(w)\mid\Xi_1=\xi]\).
In the symmetrization over \(\sigma\in\mathfrak S(\{1,\ldots,p\})\), vertex
\(1\) occupies coordinate block \(a\) of the kernel argument in exactly
\((p-1)!\) of the \(p!\) permutations, for each \(a=1,\ldots,p\). Conditioning
on \(\Xi_1=\xi=(x,u)\) and integrating out the other \(p-1\) vertices therefore
gives
\[
        \phi^0_{N,S,1}(\xi;w)
        =
        \frac{1}{p f_W(w)}
        \sum_{a=1}^p
        E\left[
        K_h\{w-X^{(a)}(x)\}
        \left\{
        \mu_S\bigl(\Xi^{(a)}(\xi)\bigr)
        -
        P\bigl(S,X^{(a)}(x)\bigr)
        \right\}
        \right]
        =
        \ell_{N,S}(\xi;w),
\]
the first-order projection defined in \eqref{eq:ell-explicit}. Hence the
H\'ajek projection of the U-statistic is
\begin{equation}
\label{eq:Hajek-projection}
        U_{N,S,1}(w)
        =
        \frac pN
        \sum_{i=1}^N
        \ell_{N,S}(\Xi_i;w).
\end{equation}

For \(k\geq2\), the M\"obius representation makes every \(\phi^0_{N,S,k}\) degenerate. To see this, condition on
\(\Xi_1,\ldots,\Xi_{k-1}\) and pair every
\(A\subseteq\{1,\ldots,k-1\}\) with \(A\cup\{k\}\). Iterated expectations give
\[
        E\left[
        \psi_{N,S,A\cup\{k\}}
        (\Xi_{A\cup\{k\}};w)
        \mid
        \Xi_1,\ldots,\Xi_{k-1}
        \right]
      =
        \psi_{N,S,A}(\Xi_A;w),
\]
while the two M\"obius coefficients have opposite signs. Therefore,
\[
        E\left[
        \phi^0_{N,S,k}(\Xi_1,\ldots,\Xi_k;w)
        \mid
        \Xi_1,\ldots,\Xi_{k-1}
        \right]
        =
        0.
\]
Symmetry gives the same result after conditioning on any \(k-1\) arguments.

I now derive the orders of the higher-order projections to show that
\begin{equation}
\label{eq:hk-L2}
        E\int
        \phi^0_{N,S,k}(\Xi_1,\ldots,\Xi_k;w)^2
        \pi(w)\,dw
        \leq
        Ch_N^{-kd_X},
        \qquad k=1,\ldots,p.
\end{equation}
Since \(\phi^0_{N,S,k}\) is a signed sum of the \(\psi_{N,S,A}\) with
\(A\subseteq\{1,\ldots,k\}\) by \eqref{eq:Mobius-kernel}, and
\(h_N^{-rd_X}\le h_N^{-kd_X}\) for \(r\le k\) once \(h_N\le1\), it suffices to
show that
\begin{equation}
\label{eq:psi-L2}
        E\int
        \psi_{N,S,A}(\Xi_A;w)^2
        \pi(w)\,dw
        \le
        C\,h_N^{-|A|d_X}
        \qquad\text{for every }A\subseteq\{1,\ldots,p\}.
\end{equation}
For
\(B\subseteq\{1,\ldots,p\}\), write \(z=(z_1,\ldots,z_p)\in\mathbb R^{d_W}\)
with blocks \(z_a\in\mathbb R^{d_X}\), let \(z_B:=(z_a)_{a\in B}\), and define
the partially integrated kernel envelope
\[
        \overline K_B(z_B)
        :=
        \int
        |K(z)|
        \,dz_{B^c},
\]
the integral of \(|K|\) over the blocks not in \(B\). By
Assumption~\ref{aspt:KernelA}, \(\sup_{z_B}\overline K_B(z_B)\le C_2\) and
\(\int\overline K_B(z_B)\,dz_B=\int|K(z)|\,dz\le C_1\); hence
\(\|\overline K_B\|_{L^2}^2\le C_1C_2\).
Fix \(A\subseteq\{1,\dots,p\}\) with \(|A|=r\ge1\), and write its conditioned arguments as
\(\xi_1,\dots,\xi_r\) with \(\xi_j=(x_j,u_j)\). In the definition of
\(\psi_{N,S,A}\), each permutation in the symmetrization of
\(\overline \Upsilon^0_{N,\{1,\dots,p\}}\) places the \(r\) conditioned
vertices into \(r\) of the \(p\) coordinate blocks of \(w=(w_1,\dots,w_p)\), and
the conditional expectation integrates the remaining \(p-r\) vertices against
the kernel. Permutations that place the conditioned vertices in the same blocks
in the same order give the same conditional expectation, so
\(\psi_{N,S,A}(\xi_A;w)\) is an average of at most \(p!/(p-r)!\) distinct
terms, one for each placement \(\tau\) of the conditioned vertices into
distinct blocks; write \(B_\tau\) for the set of blocks used, \(w_{B_\tau}\)
for those blocks of \(w\), and \(x_{B_\tau}\) for the conditioned covariates
placed in them. For \(w\in\mathcal W_\pi\), since \(|\mu_S-P|\le1\) and
\(f_W(w)^{-1}\le\underline f^{-1}\) by Assumption~\ref{ass:HM-complete}(i),
I bound the densities of the \(p-r\) free vertices by \(B_1\)
by Assumption~\ref{aspt:Model}, and substitute
\(z_{B_\tau^c}=(w_{B_\tau^c}-x_{B_\tau^c})/h_N\) for change of variables, which gives
\[
        |\psi_{N,S,A}(\xi_A;w)|
        \le
        C\,h_N^{-d_W}\,h_N^{(p-r)d_X}
        \sum_{\tau}
        \overline K_{B_\tau}\!\left(\frac{w_{B_\tau}-x_{B_\tau}}{h_N}\right)
        =
        C\,h_N^{-rd_X}
        \sum_{\tau}
        \overline K_{B_\tau}\!\left(\frac{w_{B_\tau}-x_{B_\tau}}{h_N}\right),
\]
where \(h_N^{-d_W}\) is the factor in \(K_h\) and \(h_N^{(p-r)d_X}\) the
Jacobian of the substitution.
Squaring, using \((\sum_\tau a_\tau)^2\le M\sum_\tau a_\tau^2\) with
\(M\le p!/(p-r)!\), and integrating against \(\pi\) gives, for every
\(\xi_A\),
\[
\begin{aligned}
        \int\psi_{N,S,A}(\xi_A;w)^2\pi(w)\,dw
        &\le
        C\,h_N^{-2rd_X}
        \sum_{\tau}
        \int
        \overline K_{B_\tau}\!\left(\frac{w_{B_\tau}-x_{B_\tau}}{h_N}\right)^2
        \pi(w)\,dw                                                        \\
        &\le
        C\,h_N^{-2rd_X}
        \sum_{\tau}
        \int
        \overline K_{B_\tau}\!\left(\frac{w_{B_\tau}-x_{B_\tau}}{h_N}\right)^2
        dw_{B_\tau}                                                       \\
        &=
        C\,h_N^{-2rd_X}\cdot h_N^{rd_X}
        \sum_{\tau}
        \|\overline K_{B_\tau}\|_{L^2}^2
        \le
        C\,h_N^{-rd_X},
\end{aligned}
\]
where the second inequality integrates the \(p-r\) blocks of \(w\) outside
\(B_\tau\) against \(\pi\), which is bounded with compact support, and the
equality substitutes \(u=(w_{B_\tau}-x_{B_\tau})/h_N\) in the \(r\) blocks of
\(B_\tau\). The bound does not depend on \(\xi_A\), so
\[
        E\int
        \psi_{N,S,A}(\Xi_A;w)^2
        \pi(w)\,dw
        \le
        C\,h_N^{-rd_X},
\]
which is \eqref{eq:psi-L2}, and hence \eqref{eq:hk-L2}.

Next, terms indexed by distinct \(\mathbf q_k\neq\mathbf r_k\) in
\(\mathcal C_{k,N}\) are uncorrelated. Pick \(i\in\mathbf q_k\setminus\mathbf r_k\)
and condition on \(\{\Xi_j:j\neq i\}\): the term indexed by \(\mathbf r_k\) is
a function of the conditioning variables, while the term indexed by
\(\mathbf q_k\) has conditional mean zero, because \(\phi^0_{N,S,k}\) is
canonical and \(\Xi_i\) is independent of the other vertex variables.
Consequently,
\begin{equation*}
        E\int U_{N,S,k}(w)^2\pi(w)\,dw
        =
        \frac{\binom pk^2}{\binom Nk}
        E\int
        \phi^0_{N,S,k}(\Xi_1,\ldots,\Xi_k;w)^2
        \pi(w)\,dw
        \leq
        CN^{-k}h_N^{-kd_X}.
\end{equation*}
In particular, for \(k=1\),
\(Nh_N^{d_X/2}\int U_{N,S,1}(w)^2\pi(w)\,dw=O_p(h_N^{-d_X/2})\).
For \(k\geq2\), orthogonality across Hoeffding orders, which follows by the
same argument, implies
\[
        E\int
        \left\{
        \sum_{k=2}^pU_{N,S,k}(w)
        \right\}^2
        \pi(w)\,dw
        \leq
        C\sum_{k=2}^pN^{-k}h_N^{-kd_X}.
\]

For the edge-shock remainder, conditional independence implies that its
conditional covariance is zero for two \(p\)-subsets sharing at most one
vertex, because their induced subgraphs contain no common dyadic shock. For a
pair sharing \(t\geq2\) vertices, partial kernel integration gives a bound
\(Ch_N^{-td_X}\), and there are \(O(N^{2p-t})\) such pairs. Hence
\[
        E\int V_{N,S}^{e}(w)^2\pi(w)\,dw
        \leq
        C\sum_{t=2}^pN^{-t}h_N^{-td_X}.
\]

Set
$R_{N,S}^{H}(w)
        :=
        \sum_{k=2}^pU_{N,S,k}(w)+V_{N,S}^{e}(w),$
so that \(U_{N1}=U_{N,S,1}+R_{N,S}^{H}\).
Since \(\sum_{k=2}^pU_{N,S,k}\) is a function of \(\Xi_1,\ldots,\Xi_N\) alone
whereas
\(E[V_{N,S}^{e}\mid\Xi_1,\ldots,\Xi_N]=0\), the two parts of \(R_{N,S}^{H}\) are orthogonal. Therefore,
\begin{equation*}
        E\left[
        Nh_N^{d_X/2}
        \int R_{N,S}^{H}(w)^2\pi(w)\,dw
        \right]
        \leq
        C\sum_{k=2}^p
        N^{1-k}h_N^{-(k-1/2)d_X}=
        O\left(\frac1{Nh_N^{3d_X/2}}\right).
\end{equation*}

Thus
\begin{equation}
\label{eq:Hoeffding-remainder-square}
        Nh_N^{d_X/2}
        \int R_{N,S}^{H}(w)^2\pi(w)\,dw
        =
        O_p\left(\frac1{Nh_N^{3d_X/2}}\right)
        =
        o_p(1).
\end{equation}
By the Cauchy--Schwarz inequality and the bounds on \(U_{N,S,1}\) and
\(R_{N,S}^{H}\) above,
\[
\begin{aligned}
        &\left|
        Nh_N^{d_X/2}
        \int
        U_{N,S,1}(w)R_{N,S}^{H}(w)
        \pi(w)\,dw
        \right|                                         \\
        &\qquad\leq
        \left[
        Nh_N^{d_X/2}
        \int U_{N,S,1}(w)^2\pi(w)\,dw
        \right]^{1/2}
        \left[
        Nh_N^{d_X/2}
        \int R_{N,S}^{H}(w)^2\pi(w)\,dw
        \right]^{1/2}                                   \\
        &\qquad=
        O_p(h_N^{-d_X/4})
        O_p\left(
        N^{-1/2}h_N^{-3d_X/4}
        \right)                                         \\
        &\qquad=
        O_p\left(
        N^{-1/2}h_N^{-d_X}
        \right)
        =
        O_p\left(
        \frac1{\sqrt{Nh_N^{2d_X}}}
        \right)
        =
        o_p(1).
\end{aligned}
\]
Thus,
\begin{equation}
\label{eq:projection-reduction}
        Nh_N^{d_X/2}
        \int U_{N1}(w)^2\pi(w)\,dw
        =
        Nh_N^{d_X/2}
        \int U_{N,S,1}(w)^2\pi(w)\,dw
        +
        o_p(1).
\end{equation}

\medskip
\noindent
\textit{Step 5: Diagonal and off-diagonal decomposition.}

Substituting \eqref{eq:Hajek-projection} into
\eqref{eq:projection-reduction} and separating the diagonal from the
off-diagonal terms gives
\[
        Nh_N^{d_X/2}
        \int U_{N,S,1}(w)^2\pi(w)\,dw
        =
        B_{N,S}+W_{N,S},
\]
where, with \(c_N:=2p^2h_N^{d_X/2}/N\),
\[
        B_{N,S}
        :=
        \frac{c_N}{2}
        \sum_{i=1}^N
        \int
        \ell_{N,S}(\Xi_i;w)^2
        \pi(w)\,dw,
\]
and
\[
        W_{N,S}
        :=
        c_N
        \sum_{1\leq i<j\leq N}
        \int
        \ell_{N,S}(\Xi_i;w)
        \ell_{N,S}(\Xi_j;w)
        \pi(w)\,dw.
\]

By the definition of \(\Gamma_{N,S}\),
\begin{equation*}
        E[B_{N,S}]
        =
        p^2h_N^{d_X/2}
        \int
        E[\ell_{N,S}(\Xi_1;w)^2]
        \pi(w)\,dw
        =
        h_N^{-d_X/2}
        \int
        \Gamma_{N,S}(w,w)\pi(w)\,dw
        =
        b_{N,S}.
\end{equation*}

To bound \(\var{B_{N,S}}\), I first bound \(\ell_{N,S}\) pointwise. For \(w\in\mathcal W_\pi\), since \(|\mu_S-P|\le1\)
and \(f_W(w)\ge\underline f\) by Assumption~\ref{ass:HM-complete}(i), the
definition \eqref{eq:ell-explicit} gives
\[
\begin{aligned}
        |\ell_{N,S}(\Xi_i;w)|
        &\le
        \frac{1}{p\underline f}
        \sum_{a=1}^p
        E\big|K_h\{w-X^{(a)}(X_i)\}\big|                                  \\
        &\le
        \frac{B_1^{\,p-1}}{p\underline f}\,h_N^{-d_W}
        \sum_{a=1}^p
        \int
        \left|K\!\left(\frac{w-x^{(a)}}{h_N}\right)\right|
        dx_{\{a\}^c}                                                      \\
        &=
        \frac{B_1^{\,p-1}}{p\underline f}\,h_N^{-d_W}\,h_N^{(p-1)d_X}
        \sum_{a=1}^p
        \overline K_{\{a\}}\!\left(\frac{w_a-X_i}{h_N}\right)
        \le
        Ch_N^{-d_X}
        \sum_{a=1}^p
        \overline K_{\{a\}}\!\left(\frac{w_a-X_i}{h_N}\right),
\end{aligned}
\]
where \(x^{(a)}\) has \(X_i\) in block \(a\) and the integration variables
\(x_b\), \(b\neq a\), in the other blocks; the second inequality bounds the
densities of the \(p-1\) free vertices by \(B_1\)
(Assumption~\ref{aspt:Model}), and the equality substitutes
\(u_b=(w_b-x_b)/h_N\) for \(b\neq a\), with Jacobian \(h_N^{(p-1)d_X}\).
Then,
\[
\begin{aligned}
        \int
        \ell_{N,S}(\Xi_i;w)^2
        \pi(w)\,dw
        &\leq
        Cp\,h_N^{-2d_X}
        \sum_{a=1}^p
        \int
        \overline K_{\{a\}}\!\left(\frac{w_a-X_i}{h_N}\right)^2
        \pi(w)\,dw                                                        \\
        &\leq
        C\,h_N^{-2d_X}
        \sum_{a=1}^p
        \int
        \overline K_{\{a\}}\!\left(\frac{w_a-X_i}{h_N}\right)^2
        dw_a                                                              \\
        &=
        C\,h_N^{-2d_X}\cdot h_N^{d_X}
        \sum_{a=1}^p
        \|\overline K_{\{a\}}\|_{L^2}^2
        \leq
        Ch_N^{-d_X},
\end{aligned}
\]
where the second inequality integrates the \(p-1\) blocks of \(w\) other than
\(w_a\) against \(\pi\), which is bounded with compact support; the equality
substitutes \(u=(w_a-X_i)/h_N\) in block \(a\); and the last inequality uses
\(\|\overline K_{\{a\}}\|_{L^2}^2\le C_1C_2\) from Step~4.
Since this bound holds for every realization of \(\Xi_i\), squaring it and
taking expectations gives
\[
        E\left[
        \left\{
        \int
        \ell_{N,S}(\Xi_i;w)^2
        \pi(w)\,dw
        \right\}^2
        \right]
        \leq
        Ch_N^{-2d_X}.
\]
Since the summands in \(B_{N,S}\) are i.i.d.,
\[
        \var{B_{N,S}}
        \leq
        \frac{p^4h_N^{d_X}}{N}
        Ch_N^{-2d_X}
        =
        O\left(\frac1{Nh_N^{d_X}}\right)
        =
        o(1).
\]
Therefore, $B_{N,S}=b_{N,S}+o_p(1).$

\medskip
\noindent
\textit{Step 6: de Jong's CLT for the off-diagonal quadratic term.}

For \(i<j\), define
\[
        W_{N,ij,S}
        :=
        c_N
        \int
        \ell_{N,S}(\Xi_i;w)\ell_{N,S}(\Xi_j;w)\pi(w)\,dw.
\]
Then \(W_{N,S}=\sum_{i<j}W_{N,ij,S}\). I apply the central limit theorem for clean quadratic forms of \textcite{de_jong_central_1987} by verifying its three conditions.

First, the array is clean, that is,
\(E[W_{N,ij,S}\mid\Xi_i]=E[W_{N,ij,S}\mid\Xi_j]=0\) for \(i\neq j\). Indeed,
since \(\Xi_j\) is independent of \(\Xi_i\) and \(E[\ell_{N,S}(\Xi_j;w)]=0\)
for every \(w\) by \eqref{eq:ell-explicit},
\[
        E[W_{N,ij,S}\mid\Xi_i]
        =
        c_N
        \int
        \ell_{N,S}(\Xi_i;w)\,E[\ell_{N,S}(\Xi_j;w)]\,
        \pi(w)\,dw
        =
        0,
\]
and the same argument gives \(E[W_{N,ij,S}\mid\Xi_j]=0\).

Second, let \(s_{N,S}^2:=\var{W_{N,S}}\). If two unordered index pairs
are disjoint, the corresponding \(W\)'s are independent; if they share one
index, cleanness gives zero covariance by conditioning on that index. Hence
\[
        s_{N,S}^2
        =
        \binom N2E[W_{N,12,S}^2].
\]
Independence of \(\Xi_1\) and \(\Xi_2\) gives
\[
\begin{aligned}
        E[W_{N,12,S}^2]
        &=
        c_N^2
        \int\!\!\int
        E[\ell_{N,S}(\Xi_1;w)\ell_{N,S}(\Xi_1;z)]^2
        \pi(w)\pi(z)\,dw\,dz                            \\
        &=
        \frac{4h_N^{-d_X}}{N^2}
        \int\!\!\int
        \Gamma_{N,S}(w,z)^2
        \pi(w)\pi(z)\,dw\,dz.
\end{aligned}
\]
Thus
\[
        s_{N,S}^2
        =
        \frac{N-1}{N}\mathcal V_{N,S}
        \to
        \mathcal V_S.
\]

For every \(i\),
\[
        \sum_{j\neq i}E[W_{N,ij,S}^2]
        =
        (N-1)E[W_{N,12,S}^2]
        =
        \frac2N s_{N,S}^2.
\]
Therefore de Jong's row-negligibility condition holds:
\[
        \frac1{s_{N,S}^2}
        \max_i
        \sum_{j\neq i}E[W_{N,ij,S}^2]
        =
        \frac2N
        \to0.
\]

Finally, I verify the fourth-moment condition. Substituting the pointwise
bound on \(\ell_{N,S}\) from Step~5 gives
\[
        \left|
        \int
        \ell_{N,S}(\Xi_i;w)\ell_{N,S}(\Xi_j;w)\pi(w)\,dw
        \right|
        \leq
        Ch_N^{-2d_X}
        \sum_{a,b=1}^p
        \int
        \overline K_{\{a\}}\!\left(\frac{w_a-X_i}{h_N}\right)
        \overline K_{\{b\}}\!\left(\frac{w_b-X_j}{h_N}\right)
        \pi(w)\,dw.
\]
For \(a\neq b\), the integrand depends on \(w\) through two different blocks;
integrating the other blocks against \(\pi\) and substituting
\(u_a=(w_a-X_i)/h_N\) and \(u_b=(w_b-X_j)/h_N\) gives a Jacobian
\(h_N^{2d_X}\) and leaves
\(\|\overline K_{\{a\}}\|_{L^1}\|\overline K_{\{b\}}\|_{L^1}\le C_1^2\), so
these terms are bounded by a constant. For \(a=b\), the same substitution in
block \(a\) alone gives a Jacobian \(h_N^{d_X}\) and leaves
\(\kappa_a\bigl((X_i-X_j)/h_N\bigr)\), where
\[
        \kappa_a(v)
        :=
        \int
        \overline K_{\{a\}}(u)\,\overline K_{\{a\}}(u+v)\,du
\]
is bounded by \(C_1C_2\) and integrable with
\(\int\kappa_a=\|\overline K_{\{a\}}\|_{L^1}^2\le C_1^2\). With
\(\kappa:=\sum_{a=1}^p\kappa_a\), therefore,
\begin{equation}
\label{eq:inner-product-envelope}
        \left|
        \int
        \ell_{N,S}(\Xi_i;w)\ell_{N,S}(\Xi_j;w)\pi(w)\,dw
        \right|
        \leq
        C\left[
        1+
        h_N^{-d_X}
        \kappa\left(
        \frac{X_i-X_j}{h_N}
        \right)
        \right].
\end{equation}

For \(r\geq1\) and \(j\neq1\), the substitution \(X_j=X_1-h_Nu\) gives,
uniformly in \(X_1\),
\[
        E\left[
        \kappa\left(\frac{X_1-X_j}{h_N}\right)^r
        \,\middle|\,X_1
        \right]
        =
        h_N^{d_X}
        \int\kappa(u)^r f(X_1-h_Nu)\,du
        \le
        B_1\|\kappa\|_\infty^{\,r-1}\|\kappa\|_{L^1}\,h_N^{d_X},
\]
so that, by \((1+x)^r\le2^{r-1}(1+x^r)\),
$E\left[
        \left\{1+h_N^{-d_X}\kappa\left(\frac{X_1-X_j}{h_N}\right)\right\}^r
        \,\middle|\,X_1
        \right]
        =
        O\bigl(h_N^{-(r-1)d_X}\bigr)$ uniformly in $X_1$.
Combined with \eqref{eq:inner-product-envelope}, this gives
\begin{equation}
\label{eq:r-moment}
        E\left[
        \left\{
        \int\ell_{N,S}(\Xi_1;w)\ell_{N,S}(\Xi_2;w)\pi(w)\,dw
        \right\}^r
        \right]
        =
        O\bigl(h_N^{-(r-1)d_X}\bigr),
\end{equation}
in particular \(O(h_N^{-d_X})\) for \(r=2\) and \(O(h_N^{-3d_X})\) for
\(r=4\).

Conditional on \(X_1\), the factors involving \(X_2\) and \(X_3\) are
independent, so \eqref{eq:inner-product-envelope} and the conditional bound
with \(r=2\), applied to each factor, give
\begin{equation}
\label{eq:three-vertex-moment}
\begin{aligned}
        &E\left[
        \left\{
        \int\ell_{N,S}(\Xi_1;w)\ell_{N,S}(\Xi_2;w)\pi(w)\,dw
        \right\}^2
        \left\{
        \int\ell_{N,S}(\Xi_1;w)\ell_{N,S}(\Xi_3;w)\pi(w)\,dw
        \right\}^2
        \right]                                                           \\
        &\qquad=
        O(h_N^{-d_X})\cdot O(h_N^{-d_X})
        =
        O(h_N^{-2d_X}).
\end{aligned}
\end{equation}
For the four-cycle, \eqref{eq:inner-product-envelope} bounds the product of
the four inner products by
\[
        C\prod_{(i,j)\in\{(1,2),(2,3),(3,4),(4,1)\}}
        \left\{1+h_N^{-d_X}\kappa\!\left(\frac{X_i-X_j}{h_N}\right)\right\}.
\]
Expanding the product, the term with all four localized factors is, after the
substitution \(X_1=v\) and \(X_i=v+h_Nu_i\) for \(i=2,3,4\) with Jacobian
\(h_N^{3d_X}\),
\[
\begin{aligned}
        &h_N^{-4d_X}
        E\left[
        \kappa\!\left(\frac{X_1-X_2}{h_N}\right)
        \kappa\!\left(\frac{X_2-X_3}{h_N}\right)
        \kappa\!\left(\frac{X_3-X_4}{h_N}\right)
        \kappa\!\left(\frac{X_4-X_1}{h_N}\right)
        \right]                                                           \\
        &\qquad=
        h_N^{-d_X}
        \int
        \kappa(-u_2)\kappa(u_2-u_3)\kappa(u_3-u_4)\kappa(u_4)\,
        f(v)\prod_{i=2}^4f(v+h_Nu_i)
        \,du_2\,du_3\,du_4\,dv
        =
        O(h_N^{-d_X}),
\end{aligned}
\]
since \(f\le B_1\) and \(\kappa\) is bounded and integrable. Terms with fewer
localized factors are \(O(1)\): their edges contain no cycle, so each
localized factor can be substituted separately with Jacobian \(h_N^{d_X}\).
Hence
\begin{equation}
\label{eq:four-cycle-moment}
        \left|
        E\left[
        \prod_{(i,j)\in\{(1,2),(2,3),(3,4),(4,1)\}}
        \int
        \ell_{N,S}(\Xi_i;w)\ell_{N,S}(\Xi_j;w)\pi(w)\,dw
        \right]
        \right|
        =
        O(h_N^{-d_X}).
\end{equation}

I now expand the fourth moment
\[
        E[W_{N,S}^4]
        =
        \sum_{e_1,e_2,e_3,e_4}
        E[W_{e_1}W_{e_2}W_{e_3}W_{e_4}],
\]
where each \(e_r\) ranges over the unordered vertex pairs and
\(W_e:=W_{N,ij,S}\) for \(e=\{i,j\}\) with \(i<j\). View
\((e_1,\ldots,e_4)\) as a multigraph on the vertices it covers. If some vertex
lies in exactly one of the four edges, say \(e_r=\{i,j\}\) with \(j\) in no
other edge, then, conditioning on \(\{\Xi_m:m\neq j\}\), the other three
factors are functions of the conditioning variables while
\(E[W_{N,ij,S}\mid\{\Xi_m:m\neq j\}]=E[W_{N,ij,S}\mid\Xi_i]=0\) by cleanness,
so the expectation vanishes. Hence only multigraphs in which every vertex has
degree at least two contribute.

Up to relabeling, these are of five types: (i) all four edges equal, with
\(O(N^2)\) configurations; (ii) two distinct doubled edges sharing one vertex,
with \(O(N^3)\); (iii) a triangle with one doubled edge, with \(O(N^3)\);
(iv) a four-cycle, with \(O(N^4)\); and (v) two distinct disjoint doubled
edges, the Gaussian pairings. Since
\(W_{N,ij,S}=c_N\int\ell_{N,S}(\Xi_i;w)\ell_{N,S}(\Xi_j;w)\pi(w)\,dw\) with
\(c_N^4=16p^8h_N^{2d_X}/N^4\), the moment bounds \eqref{eq:r-moment} with
\(r=4\), \eqref{eq:three-vertex-moment}, and \eqref{eq:four-cycle-moment}
give
\[
\begin{aligned}
        G_{I}&=O(N^2)\,E[W_{N,12,S}^4]
        =O(N^2)\,c_N^4\,O(h_N^{-3d_X})
        =O\bigl(N^{-2}h_N^{-d_X}\bigr),\\
        G_{II}&=O(N^3)\,E[W_{N,12,S}^2W_{N,13,S}^2]
        =O(N^3)\,c_N^4\,O(h_N^{-2d_X})
        =O(N^{-1}),\\
        G_{III}&=O(N^3)\,\bigl|E[W_{N,12,S}^2W_{N,13,S}W_{N,23,S}]\bigr|\\
        &\le
        O(N^3)\,
        \bigl(E[W_{N,12,S}^2W_{N,13,S}^2]\bigr)^{1/2}
        \bigl(E[W_{N,12,S}^2W_{N,23,S}^2]\bigr)^{1/2}\\
        &\le O(N^3)\,c_N^4\,O(h_N^{-2d_X})
        =O(N^{-1}),\\
        G_{IV}&=O(N^4)\,\bigl|E[W_{N,12,S}W_{N,23,S}W_{N,34,S}W_{N,41,S}]\bigr|
        =O(N^4)\,c_N^4\,O(h_N^{-d_X})
        =O(h_N^{d_X}).
\end{aligned}
\]
All four are \(o(1)\), since \(h_N\to0\) and \(Nh_N^{d_X}\to\infty\).
For type (v), fix two distinct disjoint pairs \(\{i,j\}\) and \(\{k,l\}\).
There are \(\binom42=6\) ways to assign them to the positions
\(r=1,\ldots,4\), and
\(E[W_{N,ij,S}^2W_{N,kl,S}^2]=E[W_{N,ij,S}^2]\,E[W_{N,kl,S}^2]\) because
disjoint pairs involve independent vertex variables. Splitting
\((\sum_{i<j}E[W_{N,ij,S}^2])^2\) into its diagonal terms, its disjoint pairs,
and its pairs sharing one vertex, the total contribution of type (v) is
\begin{align*}
       & 6\sum_{\{i,j\}\cap\{k,l\}=\varnothing}
        E[W_{N,ij,S}^2]\,E[W_{N,kl,S}^2]\\
        &=
        3\Bigl(\sum_{i<j}E[W_{N,ij,S}^2]\Bigr)^2
        -
        3\sum_{i<j}\bigl(E[W_{N,ij,S}^2]\bigr)^2
       -
        6\sum_{|\{i,j\}\cap\{k,l\}|=1}
        E[W_{N,ij,S}^2]\,E[W_{N,kl,S}^2]                                  \\
        &=
        3s_{N,S}^4+o(1),
\end{align*}
where the last equality uses
\(\sum_{i<j}E[W_{N,ij,S}^2]=s_{N,S}^2\) and
\(E[W_{N,ij,S}^2]=E[W_{N,12,S}^2]=2\mathcal V_{N,S}/N^2=O(N^{-2})\). The
diagonal sum is \(O(N^2\cdot N^{-4})=O(N^{-2})\), and the sum over the
\(O(N^3)\) pairs sharing one vertex is \(O(N^3\cdot N^{-4})=O(N^{-1})\).

Consequently, since \(s_{N,S}^2\to\mathcal V_S>0\),
${E[W_{N,S}^4]}/{s_{N,S}^4}
        \to3.$
De Jong's central limit theorem for clean generalized quadratic forms now gives
\begin{equation}\label{eq:Q-limit}
W_{N,S}
        \to_d
        N(0,\mathcal V_S).
\end{equation}

\medskip
\noindent
\textit{Step 7: Conclusion.}

By \eqref{eq:reduced-to-Un1}, \eqref{eq:projection-reduction},
\(B_{N,S}=b_{N,S}+o_p(1)\), and \eqref{eq:Q-limit},
$
        T_{N,S}
        =
        B_{N,S}+W_{N,S}+o_p(1)
        =
        b_{N,S}+W_{N,S}+o_p(1);
$
hence $T_{N,S}-b_{N,S}
        \to_d
        N(0,\mathcal V_S).$
\end{proof}

\section{Conclusion}\label{sec:conclusion}

This paper develops a specification test for parametric conditionally
independent dyad models of network formation. The test is based on the restrictions that a parametric model places on subgraphs. Under the null hypothesis and a nondegeneracy condition, the centered test statistic is asymptotically normal, with an explicit centering constant and variance.
The asymptotic theory accounts for the dependence among overlapping
tuples of agents in a single network, and the normalization of the statistic
depends only on the dimension of a single agent's characteristics, which
alleviates the curse of dimensionality.

\end{document}